%% file: CMB_lensing_forecasts_v5.tex
\documentclass[a4paper,11pt]{article}
\usepackage{jcappub} 
\usepackage[T1]{fontenc} 
\usepackage{graphicx}
\usepackage{verbatim}
\usepackage{ulem}
\usepackage{color}
\usepackage{float}
\usepackage{mathrsfs}
\usepackage{amssymb}
\usepackage{amsmath}
\usepackage{amsfonts}
\usepackage{amsthm}
\usepackage{psfrag}
\usepackage{epsfig}
\usepackage{bm}
\usepackage{subfigure}%
\usepackage{lscape}
\usepackage{comment}
\usepackage{multirow}

\input{newcomm}

\def\dddot#1{\mathinner{\buildrel\vbox{\kern5pt\hbox{...}}\over{#1}}}

\def\be{\begin{equation}}
\def\ee{\end{equation}}
\def\bq{\begin{eqnarray}}
\def\eq{\end{eqnarray}}
\def\beq{\begin{eqnarray*}}
\def\eeq{\end{eqnarray*}}

\title{\boldmath CMB lensing forecasts for constraining the primordial perturbations: adding to the CMB temperature and polarization information}
 
\author{Simon Muya Kasanda$^{1}$}
\author{and Kavilan Moodley$^{1,2}$}
\affiliation{$^1$ Astrophysics and Cosmology Research Unit \& School of
Mathematics, Statistics and Computer Science, University of KwaZulu-Natal, Durban, 4041, South Africa\\
$^2$ Kavli Institute for Theoretical Physics, University of California, Santa Barbara, USA}

\emailAdd{simon.muya.kasanda@gmail.com}
\emailAdd{moodleyk41@ukzn.ac.za}

\abstract{We forecast how current ({\planck}) and future ({\prism}) cosmic microwave background (CMB) experiments constrain the adiabatic mode and its admixtures with primordial isocurvature modes. The forecasts are based on measurements of the reconstructed CMB lensing potential and lensing-induced CMB B-mode polarization anisotropies in combination with the CMB temperature and E-mode polarization anisotropies. We first study the characteristic features of the CMB temperature, polarization and lensing spectra for adiabatic and isocurvature modes. We then consider how information from the CMB lensing potential and B-mode polarization induced by lensing
 can improve constraints on an admixture of adiabatic and three correlated isocurvature modes. We find that the CMB lensing spectrum improves constraints on isocurvature modes by at most 10\% for the {\planck} and {\prism} experiments. The limited improvement is a result of the low amplitude of isocurvature lensing spectra and cancellations between these spectra that render them only slightly detectable. There is a larger gain from using the lensing-induced B-mode polarization spectrum measured by {\prism}. In this case constraints on isocurvature mode amplitudes improve by as much as 40\% relative to the CMB temperature and E-mode polarization constraints. The addition of both lensing and lensing-induced B-mode polarization information constrains isocurvature mode amplitudes at the few percent level or better. In the case of admixtures of the adiabatic mode with one or two correlated isocurvature modes we find that constraints at the percent level or better are possible. We investigate the dependence of our results to various assumptions in our analysis, such as the inclusion of dark energy parameters, the CMB temperature-lensing correlation, and the presence of primordial tensor modes, and find that these assumptions do not significantly change our main results.}

\arxivnumber{1409.6869 ; { Preprint number: NSF-KITP-14-159}}

\begin{document}

\maketitle
\flushbottom

\section{Introduction}
\label{sec1}

\noindent
The temperature anisotropies in the cosmic microwave background (CMB) radiation are an important source of cosmological information. Containing imprints of the initial conditions after inflation, the large-scale anisotropies propagate to us largely unaltered and constitute a powerful probe of the early universe. The measurement of the CMB anisotropies by different experiments in temperature
 \cite{ACBAR_MISSION_2003,ACT_MISSION_2003,ARCHEOPS_MISSION_2004,BICEP2_MISSION_2014,BOOMERANG_MISSION_2006,CBI_MISSION_2001,COBE_MISSION_1990,
MAXIMA_MISSION_2006,PLANCK_MISSION_2014,QMAP_MAT_TOCO_MISSIONS_2002,SPT_MISSION_2004,WMAP_MISSION_2003} and polarization \cite{CAPMAP_first_results_2003,DASI_MISSION_2002,MAXIPOL_MISSION_2003,PIQUE_MISSION_1997,QUAD_first_results_2008,QUIET_MISSION_2012,VSA_first_results_2003}
with increasing precision have allowed us to constrain models of inflation by measuring the shape and amplitude of the primordial spectrum and placing upper limits on primordial non-gaussianity and tensor perturbations \cite{Story_etal2013,Bennett_etal2013,Hinshaw_etal2013,Planck_22_2013,Hou_etal2014,Das_etal2014}, with a recent claimed detection of primordial tensor perturbations \cite{Ade_etal2014}.\\

\noindent
Going beyond the temperature anisotropies, ongoing and upcoming CMB experiments have pushed in two main directions. The first set of experiments have targeted the large-angle CMB B-mode polarization anisotropies in order to detect primordial gravitational waves and constrain inflationary models via their tensor mode contribution \cite{Essinger-Hileman_etal2010,Filippini_etal2010,Ade_etal2014}. The second set of experiments have primarily focused on constraining the late-time universe by measuring with increased sensitivity the CMB temperature and polarization anisotropies on small scales, taking advantage of the CMB as a backlight on cosmic structures at low redshift \cite{Austermann_etal2012,Niemack_etal2010,POLARBEAR_etal2014}. However, these experiments are also sensitive to the primordial universe through direct constraints on the shape of the power spectrum via the spectral index and its running, and on primordial non-gaussianity \cite{Naess_etal2014,Sievers_etal2013}.\\

\noindent
Another aspect of the primordial perturbations that the CMB can probe is the initial conditions set up during inflation, namely, whether the perturbations were adiabatic or isocurvature in nature. The adiabatic mode (AD) is characterized by relative ratios in the number densities of species that are spatially uniform, leading to an overall curvature perturbation initially. In contrast isocurvature modes have spatially varying abundances of species that are balanced to have zero net spatial curvature initially. Depending on the species that are initially perturbed at most four different regular isocurvature modes are allowed, namely the neutrino isocurvature density (NID) and neutrino isocurvature velocity (NIV), the cold dark matter isocurvature (CI) and the baryon isocurvature (BI) modes \cite{Bucher_etal2000}. The general primordial cosmic perturbation is comprised of these five regular modes and their correlations \cite{Bucher_etal2000}.
A variety of models have been proposed that generate admixtures of adiabatic and isocurvature perturbations, including multi-field inflationary models \cite{Mollerach_1990,Polarski_Starobinsky_1994,GarciaBellido_Wands_1996,Linde_Mukhanov_1997,Langlois_1999,Gordon_etal2001,Langlois_2008}, axion models \cite{Efstathiou_Bond_1986,Bozza_etal2002} and curvaton models \cite{Lyth_Wands_2002,Moroi_Takahashi_2001,Bartolo_Liddle_2002,Moroi_Takahashi_2002,Lyth_etal2003,Dimopoulos_etal2003a,Dimopoulos_etal2003b}. Recently Iliesu {\it et al.} \cite{Iliesiu_etal2014} proposed a model that generates CI and NID modes correlated in a specific way with the adiabatic mode.\\

\noindent
Though pure isocurvature models are strongly constrained, subdominant correlations of isocurvature modes with the adiabatic mode are allowed by current data. Detecting even a small fraction of isocurvature perturbations would discard single-field inflationary models. This motivates us to adopt a more phenomenological approach in this paper and consider models with arbitrary correlations of isocurvature and adiabatic modes that are not tied to any specific underlying model. Previous studies have forecast constraints on isocurvature perturbations using CMB temperature and polarization power spectra alone \cite{Brown_etal2009,Reichardt_etal2009,Larson_etal2011}, or in combination with the matter power spectrum through galaxy redshift survey measurements \cite{Reid_etal2010,Percival_etal2010,Carbone_etal2011} and with Type Ia supernovae surveys \cite{Amanullah_etal2010}. Considering CMB temperature measurements alone, Bucher {\it et al.} \cite{Bucher_etal2001} showed that these were subject to degeneracies of isocurvature parameters with cosmological parameters in the most general case of the adiabatic mode correlated with three isocurvature modes. However, when large-angle ($\ell ~\ltsim ~100$) E-mode polarization data from {\planck} is added, then the combined data would constrain individual mode amplitudes to less than ten percent.\\

\noindent
These models have been confronted with the wealth of CMB data that has accumulated over the last decade, primarily from WMAP \cite{Peiris_etal2003,Hinshaw_etal2007,Komatsu_etal2011,Hinshaw_etal2013}, high-resolution ground-based experiments \cite{Enqvist_etal2000,Andrade_etal2005,Sievers_etal2007}, and more recently {\planck} \cite{Planck_22_2013}. Several different studies \cite{Stompor_etal1996,Langlois_Riazuelo_2000,Enqvist_etal2000,Amendola_etal2002} that have used CMB data alone, or in combination with other datasets such as the Sloan Digital Sky Survey (SDSS \cite{Dawson_etal2013}) matter power spectrum \cite{Tegmark_etal2004,Samushia_etal2013} and the SNLS supernovae survey \cite{Astier_etal2006}, have found that pure isocurvature models are ruled out but models with admixtures of adiabatic and isocurvature modes are allowed. There has been a series of papers constraining admixture models with a single isocurvature mode correlated with the adiabatic mode, using CMB data alone \cite{Valiviita_Muhonen_2003,Gordon_Malik_2004,Moodley_etal2004,Bean_etal2006,Savelainen_etal2013} or in conjunction with large-scale structure data \cite{Crotty_etal2003,Beltran_etal2004,Kurki-Suonio_etal2005,Beltran_etal2005}, with the most recent analyses constraining the isocurvature fraction to 3\% for fully correlated modes, and to 15\% for uncorrelated modes. Other studies have constrained models with two \cite{Bucher_etal2004,Dunkley_etal2005,Bean_etal2006} or three \cite{Moodley_etal2004,Bucher_etal2004,Dunkley_etal2005,Bean_etal2006} isocurvature modes correlated with the adiabatic mode, with the latest constraints permitting a maximum isocurvature fraction of 20\% and 40\% for two and three isocurvature modes, respectively.\\

\noindent
High resolution CMB experiments have opened up another avenue for constraining the primordial perturbations, through increasingly accurate measurements of the CMB lensing spectrum \cite{Das_etal2014,Story_etal2013,Planck_17_2013}. The gravitational lensing distortion of the CMB spectrum is sensitive to the amplitude and shape of the matter power spectrum, which, in the linear regime, provides a direct link to the primordial power spectrum. Moreover, the matter power spectrum inferred from CMB lensing is not affected by the biases that are inherent in galaxy redshift survey measurements of the matter power spectrum. In this work, we study the effect of combining the CMB lensing potential, reconstructed from CMB temperature and polarization spectra, with CMB temperature, E-mode and B-mode polarization spectra to forecast constraints on primordial isocurvature modes. We study how current ({\planck}) and future ({\prism}) cosmic microwave background experiments will constrain the adiabatic mode and its admixtures with primordial isocurvature perturbations, in combination with the standard set of cosmological parameters. In a forthcoming paper, we will use recent CMB temperature, polarization and lensing data to set updated constraints on models with correlated adiabatic and isocurvature modes. We note that CMB lensing information has been used previously to forecast constraints on isocurvature modes \cite{Santos_etal2012}. However, this study only considered a single CDM isocurvature mode correlated with the adiabatic mode in the context of the axion and curvaton models, whereas we consider more general correlations of one, two and three isocurvature modes. The paper also focused primarily on the {\planck} experiment, whereas we consider, in addition to {\planck}, more sensitive measurements of the B-mode polarization signal from {\prism}.\\

\noindent
This paper is structured as follows. In Section \ref{sec2} we compare and contrast the features of the CMB temperature, polarization and lensing potential spectra generated by adiabatic and isocurvature initial conditions. We consider in turn how lensing information adds to the CMB temperature, CMB temperature and E-mode polarization, and CMB temperature, E-mode and B-mode polarization measurements in constraining isocurvature modes. In Section \ref{sec3} we describe the Fisher matrix formalism that we use to forecast constraints on cosmological parameters and isocurvature mode amplitudes from CMB anisotropy and lensing spectra. In Section \ref{sec4} we forecast constraints on isocurvature modes from the {\planck} and {\prism} experiments and the impact of CMB lensing data on these constraints. We present our main results for three correlated isocurvature modes but also consider constraints on one or two correlated isocurvature modes. In this section, we also study the dependence of our results to certain assumptions in our analysis. Finally, in Section \ref{sec5} we discuss our results and conclusions, and highlight possible extensions of this work.\\

\section{CMB anisotropy and lensing spectra from primordial perturbations}
\label{sec2}

\noindent
Before deriving the anticipated constraints on the primordial perturbations that CMB anisotropy and lensing spectra can provide, it is useful to study the features of the CMB temperature, polarization and lensing spectra for adiabatic and isocurvature modes and to understand how these depend on cosmological parameters. To compute the CMB spectra presented in this section we used a modified version of CAMB \cite{CAMB_1999} with cosmological parameter values listed in Table \ref{fid_values}. The isocurvature auto-correlation spectra are normalized to have the same CMB power (defined in Section \ref{sec3.2}) as the adiabatic mode, and the cross-correlation power spectra are computed using the quadratic form relation given in \cite{Bucher_etal2002}.

\begin{figure}[t!]
\centering
\includegraphics[scale=1.3]{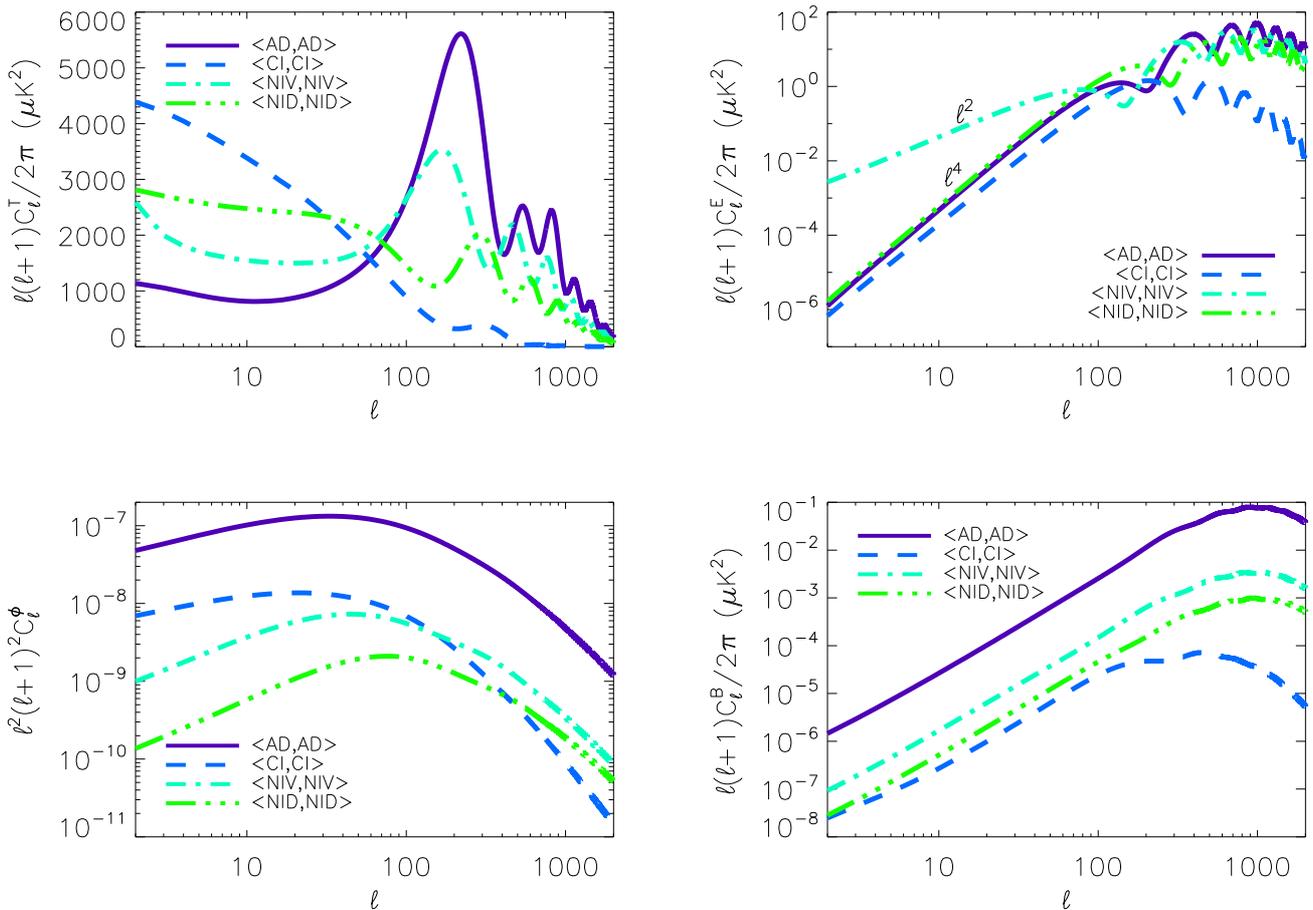}
\caption{CMB temperature (top left), E-mode polarization (top right), B-mode polarization (bottom right) and lensing (bottom left) spectra for adiabatic and isocurvature auto-correlation modes. The isocurvature observables are normalized so that their CMB temperature spectra have the same power (defined in Section \ref{sec3.2}) as the adiabatic CMB temperature spectrum. Note that the B-mode spectrum does not include tensor modes and we have switched off reionization in this figure, but have included it in our statistical treatment in the results section. For plots of the temperature and E-mode polarization cross-correlation spectra, one may refer to \cite{Bucher_etal2001}. The B-mode polarization and the lensing cross-correlation spectra are given in Figure \ref{fig_derivs_B_L}.}
\label{fig_spectra_TELB}
\end{figure}

\subsection{Temperature spectrum} \label{Section_temp_spectrum}

\noindent
The phenomenology of the CMB temperature spectrum has been studied in great detail \cite{Efstathiou_Bond_1986,Hu_Sugiyama_1994,Bucher_etal2000,Lewis_etal2000,Trotta_PhD2004,Giovannini_2005}. The features of the temperature spectrum are mainly determined by the photon density and velocity divergence, and the gravitational potential at decoupling. The photon density evolution equation in the tight-coupling regime is given by
\begin{align}
\delta_{\gamma}(k,\tau) &=A_S \, \sin{kr_s(\tau)} + A_C \cos{kr_s(\tau)} \nonumber\\
\label{deltag_bi}
&+\int_0^{\tau} (1+R(\tau'))^{1/2}\sin{[kr_s(\tau)-kr_s(\tau')]}F(k,\tau')d\tau',
\end{align}
where $A_S(k)$ and $A_C(k)$ set the initial conditions, $r_s$ is the sound horizon, $R$ is the baryon-to-photon density ratio, $\tau$ the conformal time and $F(k,\tau)$ is a source term depending on the gravitational potential and its time derivatives (see \cite{Hu_Sugiyama_1994,MuyaKasanda_etal2012} for more details). The BI, CI and NIV modes excite the $\sin{(kr_s)}$ harmonic ($A_C=0$), while the NID mode excites the $\cos{(kr_s)}$ harmonic ($A_S=0$) and the adiabatic mode is sourced purely by the gravitational driving term ($A_S=A_C=0$). The photon velocity divergence can be derived from the density using $\dot{\theta}_{\gamma}=\frac{k^2}{4}\delta_{\gamma}$. The gravitational driving term is negligible for the NIV and NID modes but becomes significant for the BI and CI modes in the matter dominated era. Figure \ref{fig_spectra_TELB} shows the CMB temperature spectra for the adiabatic and isocurvature mode auto-correlations. We note that the acoustic peak positions in the CMB spectrum follow the pattern of a harmonic series \cite{Hu_Sugiyama_1995a} that is set by the respective harmonic for a given mode. The AD, NIV and NID spectra are constant on large scales and decrease due to Silk damping on small scales \cite{Hu_Sugiyama_1994} while the BI and CI spectra steeply decrease in power on all scales due to a redistribution of power from small scales to large scales (see e.g., \cite{MuyaKasanda_etal2012}). These differing features allow one to distinguish adiabatic and isocurvature modes using the CMB temperature spectrum.\\

\subsection{E-mode polarization spectrum}
\label{sec2.2}

\noindent
The E-mode polarization of the CMB arises from a non-zero photon quadrupole at recombination due to the weakening of the photon-baryon coupling \cite{Zaldarriaga_2004}. A recent study by Galli {\it et al.} \cite{Galli_etal2014} has demonstrated that E-mode polarization measurements can constrain cosmological parameters better than CMB temperature measurements. Bucher {\it et al.} \cite{Bucher_etal2001,Bucher_etal2002} highlighted the role of large-scale E-mode polarization data in constraining the presence of primordial isocurvature modes.\\

On intermediate and small scales ($\ell\, \gtsim \,100$), the relative oscillatory behaviour in the E-mode polarization isocurvature and adiabatic spectra is similar to the temperature spectra. This is due to the phase difference between the photon density and velocity divergence, so that very little additional information is provided by E-mode polarization spectra on these scales. However, on large scales ($\ell<100$) one observes significantly different E-mode polarization spectra for adiabatic and isocurvature modes. The difference between adiabatic and isocurvature polarization spectra is studied in more detail in \cite{Moodley_MuyaKasanda_2013}. We present the relevant points here.\\

The E-mode polarization spectrum is given by
\begin{equation}
C_{\ell}^E=\frac{2}{\pi}\int k^3 d\ln{k}|\Delta_{\ell}^E(k,\tau_*)|^2 \, P(k),
\end{equation}
where $\tau_*$ is the conformal time at recombination and 
$P(k)\sim k^{-3}$ is the scale-invariant primordial power spectrum. The polarization transfer function on large-scales, $\Delta_{\ell}^E(k,\tau_*)\propto \sigma_{\gamma}(k,\tau_*) ~ \jmath_{\ell}[k(\tau_0-\tau_*)],$ is the spherical projection of the photon shear. The photon shear couples to density perturbations and is proportional to $\{k^2$, $k^2$, $k^2$, $k\}$ for the \{AD, CI, NID, NIV\} modes, respectively. Thus, on large scales, the E-mode polarization spectrum $C_{\ell}^{E}$ is proportional to $\ell^2$ for the AD, CI and NID modes, but remains constant for the NIV mode.
\footnote{Here, we have made use of the relation 
\begin{equation}
\int_{0}^{\infty}dx x^{n-2} \jmath_{\ell}^2(x)=2^{n-4}\frac{\Gamma(\ell+\frac{n}{2}-\frac{1}{2})\Gamma(3-n))}{\Gamma(\ell+\frac{5}{2}-\frac{n}{2})\Gamma^2(2-\frac{n}{2})}.
\end{equation}
}
This means that the CMB polarization spectrum for the NIV mode has more power on large scales compared to the other modes due to the larger photon velocity divergence on these scales. The turnover in the large-scale E-mode spectra occurs at an angular scale set by the photon mean free path at decoupling. The exact scale is determined by the harmonic that is excited by a given mode, as described in Section \ref{Section_temp_spectrum}. Figure \ref{fig_spectra_TELB} shows the E-mode polarization spectrum for different modes. The power law slope and turnover scale in the large-scale E-mode polarization spectra agree well with the above description.\\

\subsection{Lensing spectrum}
\label{sec2.3}
\noindent

The CMB lensing potential is determined by the initial power spectrum and the growth function of matter perturbations. The growth rate of the perturbations sets the amplitude and the turnover scale of the adiabatic and isocurvature lensing spectra. For all modes, sub-horizon fluctuations grow proportional to $\mbox{ln} ~\tau$ in the radiation dominated era and proportional to $\tau^2$ in the matter dominated era. However, super-horizon matter fluctuations grow in both the radiation era and matter era for only the adiabatic mode, resulting in a much larger amplitude of the lensing potential for the adiabatic mode compared to the isocurvature modes. Matter perturbations grow on super-horizon scales for the CI and BI modes, but only from the onset of the matter era \cite{Kodama_Sasaki_1986, Gordon_etal2001}. Matter perturbations in the NID and NIV modes only grow once the perturbations become sub-horizon.
The smaller amplitudes of the lensing power spectra for the isocurvature modes relative to the adiabatic mode are shown in Figure \ref{fig_spectra_TELB}. The amplitude of the CI lensing power spectrum is ten times smaller than its adiabatic counterpart on large scales \cite{Komatsu_etal2011} and even smaller on smaller scales, while the amplitude of the neutrino isocurvature lensing power spectra are at least two orders of magnitude smaller on large angular scales. As we will see the much smaller amplitudes of the isocurvature lensing spectra make them harder to constrain using CMB lensing data.

\subsection{Lensing-induced B-mode polarization spectrum}
\noindent

Previous studies have shown that the CMB B-mode polarization spectrum is informative in constraining the physics of the early universe (see e.g., \cite{Cabella_Kamionkowski_2004,Baumann_etal2009}). The B-mode polarization spectrum on small scales is generated by the remapping of the large-scale E-mode polarization through gravitational lensing and is given by \cite{Lewis_Challinor_2006}
\begin{equation}
C_{\ell}^{B}=\int \frac{d^2\bf{l}'}{(2\pi)^2}\left[\bf{l}'\cdot(\bf{l}-\bf{l}')\right]^2 C_{|\bf{l}-\bf{l}'|}^{\phi}C_{\bf{l}'}^{E}\sin^2{2(\phi_{\bf{l}'}-\phi_{\bf{l}})},
\end{equation}
where $\phi_{\bf l}$ is the projected gravitational potential along the line of sight. This `late' B-mode polarization, though it cannot not tell us about tensor perturbations in the early universe, can potentially constraint the primordial perturbations through its scalar contribution. This is because it depends on the existing E-mode polarization, which has a spectrum that in turn depends on the initial conditions. 
The B-mode polarization spectrum from lensing has a white spectrum on large scales $(\bf{l}\ll\bf{l}')$ \cite{Lewis_Challinor_2006} as can be seen in Figure \ref{fig_spectra_TELB}. On small scales ($\ell \geq 10000$) $C_{\ell}^{B} \propto \ell^2 C_{\ell}^{\phi}$ and on intermediate scales $C_{\ell}^{B}$ tracks $C_{\ell}^{E}$ \cite{Lewis_Challinor_2006}. This implies that there is information in the small-scale B-mode spectrum to discriminate between the initial conditions, via the information it inherits from the E-mode polarization and lensing potential spectra. The amplitude of the B-mode polarization spectrum is determined by a combination of the lensing potential and the E-mode polarization spectrum, while the turnover scale is set mainly by the E-mode polarization spectrum. \\

\section{Statistical method}
\label{sec3}
The aim of this section is to outline the formalism used to quantify the constraints on isocurvature modes based on current and next-generation datasets. We apply the Fisher matrix formalism to different generations of CMB experiments. We consider a current and a future CMB experiment that measures temperature and polarization spectra, and a reconstructed lensing potential spectrum.\\

\subsection{Fisher matrix formalism}
The accuracy with which cosmological parameters can be measured from a given dataset or experiment is conveniently quantified using the Fisher matrix formalism (see, e.g. \cite{Tegmark_etal1997}), where the Fisher matrix is given by 
\be
F_{ij} = - \left\langle \frac{\partial^2 \ln \cal{L}(\bf{x}; \bf{\theta})}{\partial \theta_i \partial \theta_j}\right\rangle.
\label{eq:FM} 
\ee 
In this formalism, the likelihood $\cal{L}(\mbox{\textbf{x}},\boldsymbol{\theta})$ of observing a set of data \textbf{x} given a model with parameters $\boldsymbol{\theta},$ depends on a set of cosmological parameters $\boldsymbol{\theta}$ that we wish to estimate. The 1-$\sigma$ errors on the parameters $\theta_i$ are given by $\sigma_i =\sqrt{F_{ii}^{-1}}$. Here \textbf{x}, the CMB power spectrum or lensing potential spectrum in this case, is modelled as an N-dimensional random variable.\\

\noindent
In the case of CMB temperature measurements alone, the Fisher matrix can be written explicitly as
\be
F_{ij}^{T} = \sum_{\ell=\ell_{min}}^{\ell_{max}^T} \frac{\partial C_{\ell}^{T}}{\partial \theta_i}\frac{\partial C_{\ell}^{T}}{\partial \theta_j}\frac{(2\ell +1)f_{sky}}{2(C_{\ell}^{T}+ {N}_{\ell}^{T})^2},
\ee
or, when E-mode polarization measurements are combined with temperature measurements, as,
\be
F_{ij}^{T+E}= \sum_{\ell=\ell_{min}}^{\ell_{max}^T}\sum_{X,Y} \frac{\partial C_{\ell}^{X}}{\partial \theta_i}\left(\mbox{Cov}_{\ell}\right)_{XY}^{-1}\frac{\partial C_{\ell}^{Y}}{\partial \theta_j}.
\ee
Here the superscripts $X,Y \in \{T,E\}$ and $\left(\mbox{Cov}_{\ell}\right)_{XY}$ is the covariance matrix between any pair of the T, E and TE power spectra, as given in \cite{Eisenstein_etal1999}. In addition we include in our Fisher matrix analysis information from the B-mode polarization spectrum,
\be
F_{ij}^{B} = \sum_{\ell=\ell_{min}}^{\ell_{max}^B} \frac{\partial C_{\ell}^{B}}{\partial \theta_i}\frac{\partial C_{\ell}^{B}}{\partial \theta_j}\frac{(2\ell +1)f_{sky}}{2(C_{\ell}^{B}+ {N}_{\ell}^{B})^2},
\ee
and from the CMB lensing potential spectrum,
\be
F_{ij}^{L} = \sum_{\ell=\ell_{min}}^{\ell_{max}^L} \frac{\partial C_{\ell}^{L}}{\partial \theta_i}\frac{\partial C_{\ell}^{L}}{\partial \theta_j}\frac{(2\ell +1)f_{sky}}{2(C_{\ell}^{L}+ {\cal N}_{\ell}^{L})^2},
\ee
where $f_{sky}$ is the fraction of sky that the experiment covers, ${N}_{\ell}$ is the beam-weighted detector noise of the experiment for temperature or polarization, and ${\cal N}_{\ell}$ is the noise arising from the quadratic estimator used in the reconstruction of the lensing signal. We use $\ell_{min}=2$ and $\ell_{max} =2000$ for all Fisher matrices in our main study but consider modifications to the value of $\ell_{min}$ in Section \ref{sec4.5}.\\

We assume that the B-mode polarization and the lensing potential Fisher matrices are independent of the CMB temperature or CMB temperature and polarization Fisher matrices. This allows us to simply sum the corresponding Fisher matrices when combining different probes. However, as pointed out in \cite{Schmittfull_etal2013}, the CMB lensing and the CMB temperature and polarization probes are correlated. We consider the impact of the correlation on our results in Section \ref{sec4.5}. We have also checked that leaving out the information from the large-scale lensing-temperature correlation, as included in previous CMB anisotropy and lensing Fisher matrix studies \cite{Perotto_etal_2006, Santos_etal2012}, has a negligible effect on our main results. We now describe the different ingredients of the Fisher matrix analysis in turn.\\

\begin{table}[t]
\begin{center}
\begin{tabular}{cccccc}
\hline
\hline
&&Fiducial & model\\
\hline
$\omega_b$& $\omega_c$& $\Omega_{\Lambda}$& $\tau$& $n_s$& $A_s$\\
\hline
0.02258& 0.11090& 0.734& 0.088& 0.963& 15.7\\
\hline
\hline
\end{tabular}
\end{center}
\caption{Parameter values for the fiducial cosmological model.}
\label{fid_values}
\end{table}

\subsection{Fiducial cosmological model and parameter derivatives}
\label{sec3.2}

The first ingredient in the Fisher matrix that we describe is the fiducial (or target) cosmological model about which the Fisher matrix is computed. We consider a spatially flat adiabatic $\Lambda$CDM model described by the following parameters: the baryon density $\omega_b$, the CDM density $\omega_c$, the cosmological constant $\Omega_{\Lambda}$, the optical depth $\tau$, the spectral index $n_s$ and the scalar amplitude $A_s,$ with their values specified in Table \ref{fid_values}. The parameter $A_s$ rescales the unit power CMB temperature spectrum to its usual amplitude via $C_{\ell}=13000\mu\mbox{K}^2\mbox{ A}_s\mbox{ }\hat{C}_{\ell}$, where $\hat{C}_{\ell}$ is the fiducial CMB temperature spectrum with unit power. We include ten additional mode amplitude parameters for the fractional contribution of adiabatic and isocurvature (CI, NID and NIV) auto-correlations and cross-correlations to the total power spectrum as described in \cite{Bucher_etal2002}. We do not include the BI mode as it produces CMB spectra that are nearly identical to the CI mode. The isocurvature spectra are fixed, up to an amplitude, by the cosmological parameters listed in Table \ref{fid_values}. Note that the Fisher matrix errors depend on the fiducial model assumed. The adiabatic fiducial model chosen here is well motivated as it provides an excellent fit to current data, and we are interested in constraining fractional contributions of isocurvature modes about the adiabatic model.\\

\begin{figure}[t!]
\centering
\includegraphics[scale=1.3]{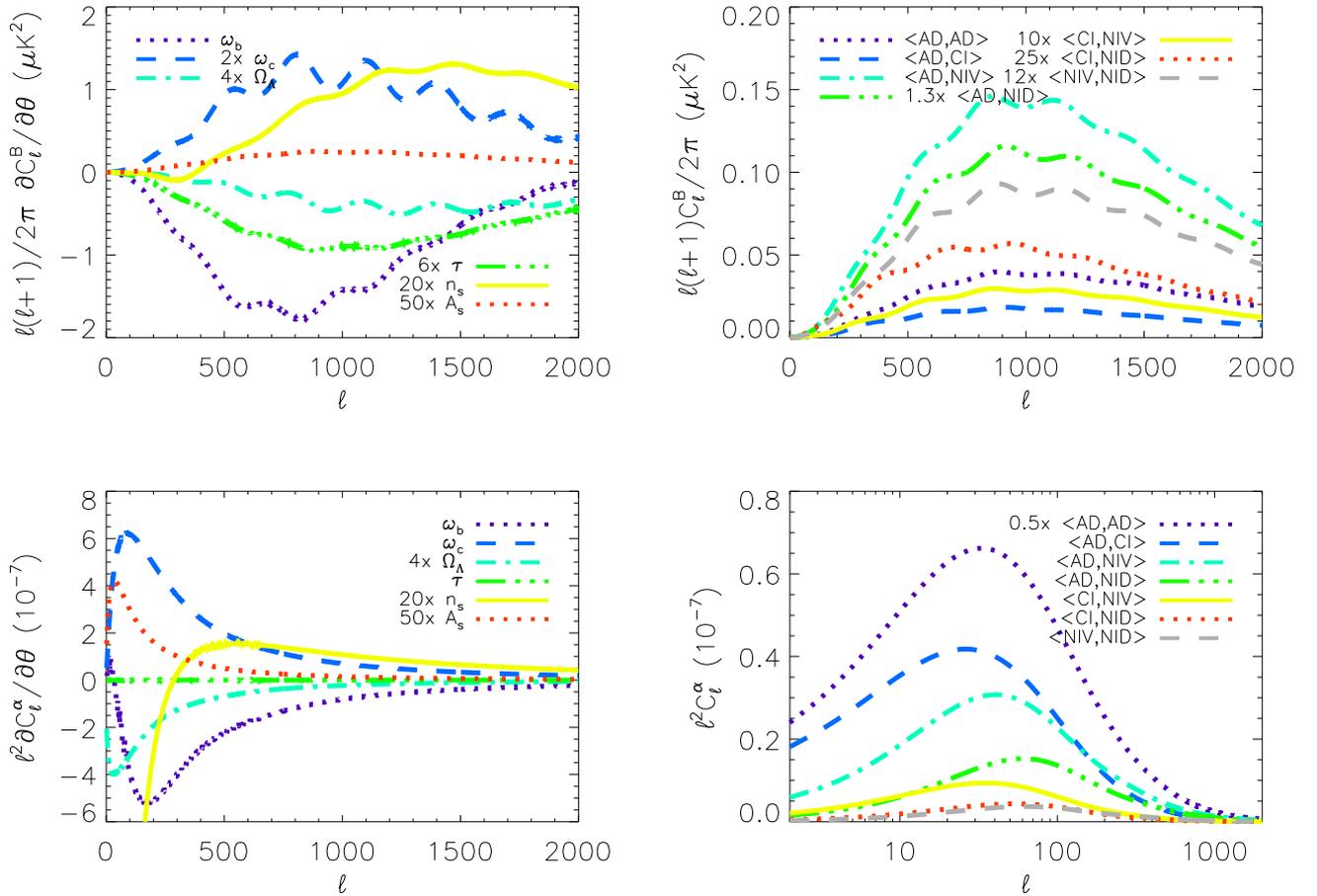}
\vskip -0.5cm 
\caption{Derivatives of the B-mode polarization and lensing spectra with respect to cosmological parameters and cross-correlation mode amplitudes. Lensing and B-mode polarization spectra for the isocurvature auto-correlation modes are shown in Figure \ref{fig_spectra_TELB}.}
\label{fig_derivs_B_L}
\end{figure}

The second ingredient in the Fisher matrix involves the derivatives of the CMB and lensing spectra with respect to the cosmological and mode amplitude parameters. The temperature and E-mode polarization spectrum derivatives have been presented in \cite{Bucher_etal2001}. In Figure \ref{fig_derivs_B_L}, we show the B-mode polarization spectrum and the lensing potential spectrum derivatives with respect to cosmological and mode amplitude parameters. For the calculation of the power spectrum derivatives with respect to the isocurvature amplitudes, we have adopted the convention in \cite{Bucher_etal2002}, where the pure isocurvature modes are normalized to have the same power i.e., $P = \sum_{\ell=2}^{2000} (2\ell+1)C_{\ell}$, in their CMB temperature spectra as the adiabatic mode. \\

\begin{table}[t]
\begin{center}
\begin{tabular}{lcccc|cccc}
\hline
\hline
&$\nu /$ GHz~$\mathbf{^a}$ & $\theta_F$~$\mathbf{^b}$ & $\Delta_T$($\mu$K)~$\mathbf{^c}$ & $\Delta_P$($\mu$K)~$\mathbf{^d}$ &$\nu /$ GHz~$\mathbf{^a}$ & $\theta_F$~$\mathbf{^b}$ & $\Delta_T$($\mu$K)~$\mathbf{^c}$ & $\Delta_P$($\mu$K)~$\mathbf{^d}$\\
\hline
\hline 
{\planck}&
143 & 7.1' & 6.0 & 11.5 & 217 & 5.0' & 13.1 & 26.8 \\
\hline
&90&5.7'&18.8&26.6&185&2.8'&7.05&9.97\\
&105&4.8'&13.8&19.6&200&2.5'&6.48&9.17\\
\raisebox{2.0ex}{{\prism}}&
135&3.8'&9.85&13.9&220&2.3'&6.26&8.85\\
&160&3.2'&7.78&11.0 &--&--&--&--\\
\hline
\hline
\end{tabular}
\end{center}
\caption{Summary of the experimental specifications for {\planck} \cite{PLANCK_MISSION_2006} and {\prism} \cite{PRISM_whitepaper2013} . $\mathbf{^a}$~Nominal central frequency of the detectors in each band. $\mathbf{^b}$~Beam full width at half maximum (FWHM). $\mathbf{^c}$~1-$\sigma$ sensitivity to intensity per frequency channel. $\mathbf{^d}$~1-$\sigma$ sensitivity to polarization per frequency channel. We assume $\Delta_P=\sqrt{2}\Delta_T$ which is a fair assumption for fully-polarized detectors. The 1-$\sigma$ sensitivities quoted are per beam.} 
\label{cmb_exp_spec}
\end{table}

\subsection{CMB experiments}
The final ingredient in the Fisher matrix analysis is the sensitivity of the CMB experiment, which depends on the detector noise and the beam width. We consider three generations of CMB experiments: a current experiment with {\planck}-temperature only \cite{PLANCK_MISSION_2006}, with data for this experiment already available publicly, and a near-term experiment with {\planck} temperature and E-mode polarization, the data for which will be released shortly. The third experiment is a future CMB polarization satellite focused on measuring the B-mode polarization signal very precisely, with experiments such as {\pixie}, {\core} and {\prism} \cite{PIXIE_whitepaper2011,CORE_whitepaper2011,PRISM_whitepaper2013} being proposed. In this paper we focus on forecasts with {\prism}. We refer to the three experiments as {\planck}-T, {\planck}-TE and {\prism}-TEB, respectively. For the CMB temperature, E-mode and B-mode measurements, the detector noise is given by 
\be
{N}_{\ell}^T=\left(\frac{\Delta_T}{T_{CMB}}\right)^2 e^{\ell(\ell+1)\theta_F^2/8\ln 2} \qquad \mbox{and} \qquad
 {N}_{\ell}^E={N}_{\ell}^B=\left(\frac{\Delta_P}{T_{CMB}}\right)^2 e^{\ell(\ell+1)\theta_F^2/8\ln 2},
\ee
where $\Delta_{T,P}$ is the noise per pixel of the detector, $\theta_F$ is the beam FWHM and we combine the noise in each frequency channel in quadrature. We follow the analysis in \cite{Albrecht_etal2006} and model the {\planck} and {\prism} datasets as CMB temperature and polarization maps of $80\%$ of the sky in two frequency bands for {\planck} and seven bands for {\prism}. The maps are considered free from any foreground contribution, assuming that the other frequency channels are used to remove them. The specifications for each experiment are given in Table \ref{cmb_exp_spec}. \\

For the lensing potential, the noise ${\cal N}_{\ell}^{L}$ is taken to be the noise of the minimum variance estimator, which is given by \cite{Hu_Okamoto_2002}
\begin{equation}
{\cal N}(\ell)=\frac{1}{\sum_{\alpha\beta}N^{-1}_{\alpha\beta}(\ell)},
\end{equation}
where $N_{\alpha\beta}$ is the noise for the estimator $\alpha\beta$ with $\{\alpha, \beta\} \in\{ TT, TE, TB, EE, EB\}$. 
We assume that each experiment uses its respective CMB temperature and polarization maps to perform the lensing map reconstruction, so that
the lensing noise is determined by the appropriate combinations of $\alpha$ and $\beta.$

\section{Results}
\label{sec4}
Our study focuses on the improvement in constraints on isocurvature perturbations when CMB lensing measurements and CMB B-mode polarization measurements are added to CMB temperature and E-mode polarization measurements. We present results for the {\planck}-T, {\planck}-TE and {\prism}-TEB experiments in Sections \ref{sec4.1}, \ref{sec4.2} and \ref{sec4.3}, respectively. In our main study we focus on the full set of three isocurvature modes correlated with the adiabatic mode, but in Section \ref{sec4.4} we present results for the case of one and two correlated isocurvature modes. In Section \ref{sec4.5} we study the sensitivity of our results to certain assumptions in our main analysis, including the sensitivity to the inclusion of dark energy parameters.\\

\subsection{CMB temperature spectrum constraints}
\label{sec4.1}

We present error forecasts on the cosmological and mode parameters for the case of a pure adiabatic mode in comparison to a model with an admixture of adiabatic and all three isocurvature modes, using {\planck} CMB temperature measurements ({\planck}-T) \cite{Planck_1_2013}. This case serves as a check of our forecasted errors compared to actual errors on adiabatic and isocurvature modes measured by {\planck} \cite{Planck_1_2013}. It is also useful for comparing to the {\planck}-TE and {\prism}-TEB forecasts.\\

The 1-$\sigma$ errors derived from the Fisher matrix for the {\planck}-T experiment are listed in Table \ref{planckt_table}. The results for the adiabatic mode are in reasonable agreement with the values derived by the {\planck} team \cite{Planck_1_2013}, though optimistic by a factor of $\sim 1.5$, as would be expected from a Fisher matrix analysis that provides minimum errors on parameters. For the adiabatic mode we observe that the addition of lensing information improves the constraints on $\omega_b$, $\omega_c$, $\Omega_{\Lambda}$ and $n_s$ marginally while the constraints on $A_s$ and $\tau$ improve by up to 40\%. This is because the lensing measurement constrains the scalar amplitude directly and reduces the impact of the $\tau$-$A_s$ degeneracy inherent in CMB temperature measurements \cite{Planck_1_2013}.\\ 

\begin{table}[t]
\begin{center}
\begin{tabular}{|l|c|c|c|c|}
\hline
& T & T+L$_T$& T & T+L$_T$\\ 
\hline
$\delta \omega_b/\omega_b$&  0.901&  0.876&  3.20&  2.82\\
\hline
$\delta \omega_c/\omega_c$&  1.82&  1.76&  5.58&  5.53\\
\hline
$\delta \Omega_{\Lambda}/\Omega_{\Lambda}$&  1.48&  1.42&  3.80&  3.76\\
\hline
$\delta \tau/\tau$&  44.6&  28.0&  54.3&  42.9\\
\hline
$\delta n_s/n_s$&  0.527&  0.520&  2.16&  2.11\\
\hline
$\delta A_s/A_s$&  7.67&  4.60&  31.6&  28.1\\
\hline
$\langle AD,CI\rangle$& -- & -- &  35.6&  32.8\\
\hline
$\langle AD,NIV\rangle$& -- & -- &  47.3&  46.5\\
\hline
$\langle AD,NID\rangle$& -- & -- &  53.1&  49.7\\
\hline
$\langle CI,CI\rangle$& -- & -- &  38.6&  38.5\\
\hline
$\langle CI,NIV\rangle$& -- & -- &  22.2&  19.9\\
\hline
$\langle CI,NID\rangle$& -- & -- &  56.0&  55.7\\
\hline
$\langle NIV,NIV\rangle$& -- & -- &  18.1&  18.0\\
\hline
$\langle NIV,NID\rangle$& -- & -- &  39.2&  37.7\\
\hline
$\langle NID,NID\rangle$& -- & -- &  16.1&  14.5\\
\hline
\end{tabular}
\end{center}
\caption{1-$\sigma$ percentage errors on cosmological parameters and isocurvature mode amplitudes for the {\planck}-T experiment.
The errors presented here and in Tables \ref{planckte_table}-\ref{planck_prism2isomodes} are computed from the inverse Fisher matrix about an adiabatic fiducial model.}
\label{planckt_table}
\end{table}

 \begin{figure}[t!]
\centering
\includegraphics[scale=1.3]{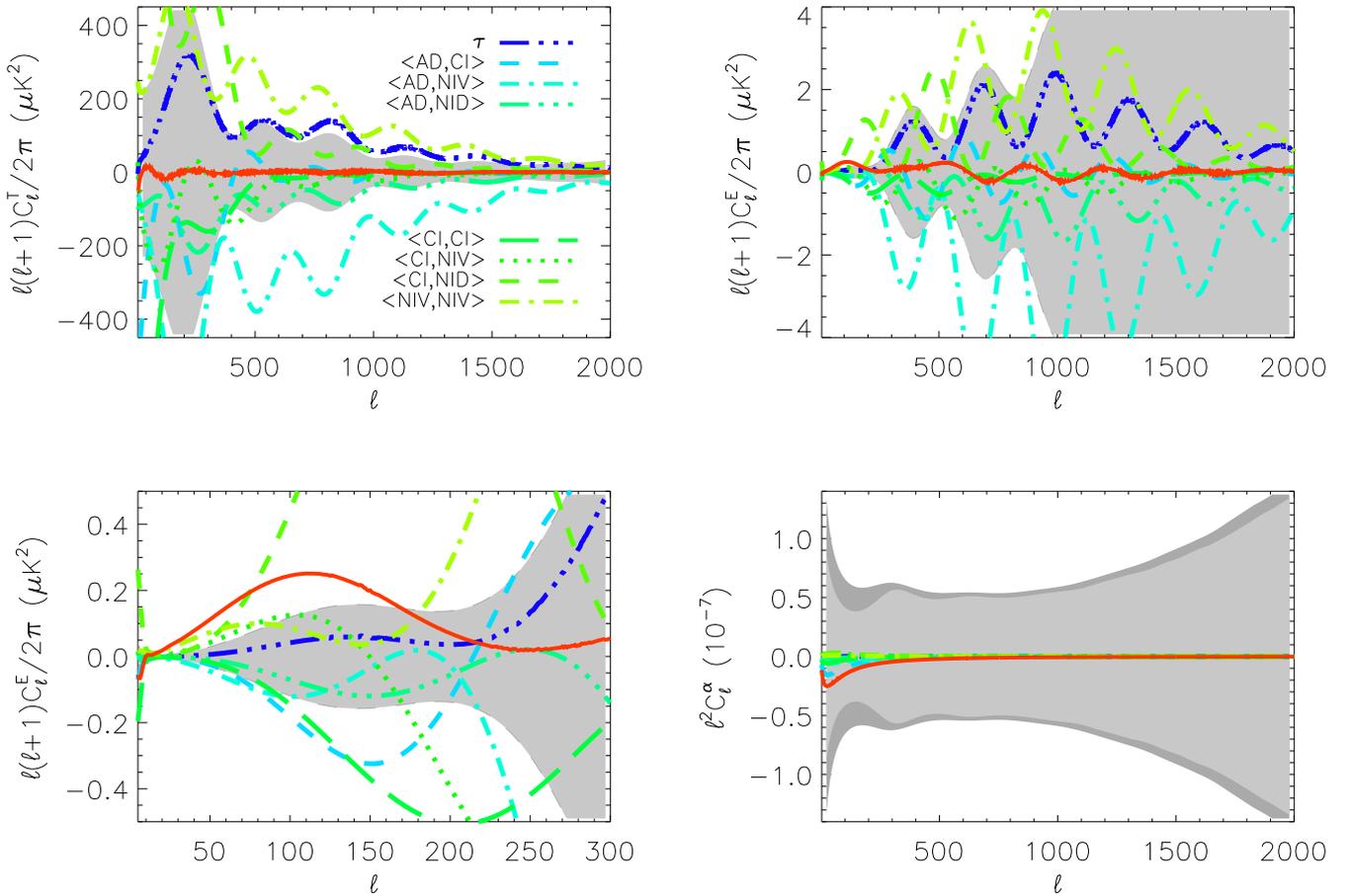}
\vskip -0.5cm 
\caption{Degeneracy breaking. The top left panel shows the cancellation between cosmological and mode amplitude parameter derivatives using the most degenerate direction of {\planck}-T measurements. The degenerate direction is obtained from a principal component analysis of the Fisher matrix. The red solid line is the sum of all contributions and is compared to the power spectrum error represented by the grey shaded region. The top right panel (all $\ell$s) and bottom left panel (low $\ell$s) show how E-mode polarization measurements constrain this degenerate direction. The bottom right panel shows how 
lensing information adds to the CMB temperature degenerate direction, with the light- and dark-grey regions representing the lensing power spectrum errors from {\planck}-TE and {\planck}-T measurements, respectively.
}
\label{fig_cancellation_planckte}
\end{figure}

When an admixture of three isocurvature modes and the adiabatic mode is allowed, the constraints on cosmological parameter degrade severely, by up to a factor of four in some cases. The isocurvature mode amplitudes are poorly constrained at the level of $50\%$. When lensing measurements are added the constraints on cosmological parameters in the admixture model improve marginally, with a more significant improvement in $\tau$ of about 20\%. The improvement in constraints on the isocurvature mode amplitudes is modest, at the level of 10\%. 
This demonstrates that {\planck} CMB lensing measurements, in which the lensing field is reconstructed from a CMB temperature map alone, do not add significantly to CMB temperature constraints on isocurvature mode amplitudes. \\

To explore this further we used a principal component analysis to determine the parameters that have significant contributions to the most degenerate direction in the {\planck}-T Fisher matrix. We compared the sum of their CMB temperature and lensing spectra, weighted appropriately with the corresponding eigenvalues, to the anticipated measurement errors for these spectra. In Figure \ref{fig_cancellation_planckte}, the top left panel shows that there are a number of isocurvature mode amplitudes that contribute to the degeneracy, together with the optical depth, $\tau$. The bottom right panel in Figure \ref{fig_cancellation_planckte} shows that the lensing measurement does not contribute significantly to reducing this degeneracy. This is primarily due to partial cancellations between the isocurvature auto-correlation and cross-correlation lensing spectra, and their low amplitude relative to the adiabatic lensing spectrum. These factors result in a combined spectrum that is smaller than the lensing measurement error. The low amplitude of the isocurvature lensing spectra relative to the adiabatic lensing spectrum results from the lack of growth of isocurvature matter perturbations relative to adiabatic matter perturbations as discussed in Section \ref{sec2.3}.

 \begin{table*}[t]
\begin{center}
\begin{tabular}{|l|c|c|c|c|}
\hline
& T~(+E) & T+L$_{TE}$~(+E) & T~(+E) & T+L$_{TE}$~(+E) \\ 
\hline
$\delta \omega_b/\omega_b$&  0.901~(0.669) &  0.873~(0.634) &  3.20~(1.05) &  2.79~(0.955) \\
\hline
$\delta \omega_c/\omega_c$&  1.82~(1.24) &  1.75~(1.07) &  5.58~(1.70) &  5.50~(1.65) \\
\hline
$\delta \Omega_{\Lambda}/\Omega_{\Lambda}$&  1.48~(1.01) &  1.41~(0.870) &  3.80~(1.23) &  3.74~(1.15) \\
\hline
$\delta \tau/\tau$&  44.6~(5.73) &  24.2~(5.67) &  54.3~(6.79) &  42.1~(6.77) \\
\hline
$\delta n_s/n_s$&  0.527~(0.391) &  0.518~(0.353) &  2.16~(0.727) &  2.10~(0.720) \\
\hline
$\delta A_s/A_s$&  7.67~(0.996) &  3.89~(0.946) &  31.6~(4.47) &  27.8~(3.87) \\
\hline
$\langle AD,CI\rangle$& -- & -- &  35.6~(12.2) &  32.5~(11.3) \\
\hline
$\langle AD,NIV\rangle$& -- & -- &  47.3~(6.16) &  46.3~(5.96) \\
\hline
$\langle AD,NID\rangle$& -- & -- &  53.1~(9.42) &  49.3~(7.87) \\
\hline
$\langle CI,CI\rangle$& -- & -- &  38.6~(6.01) &  38.5~(6.01) \\
\hline
$\langle CI,NIV\rangle$& -- & -- &  22.2~(5.59) &  19.7~(5.01) \\
\hline
$\langle CI,NID\rangle$& -- & -- &  56.0~(8.25) &  55.6~(7.80) \\
\hline
$\langle NIV,NIV\rangle$& -- & -- &  18.1~(1.63) &  18.0~(1.62) \\
\hline
$\langle NIV,NID\rangle$& -- & -- &  39.2~(5.14) &  37.6~(4.49) \\
\hline
$\langle NID,NID\rangle$& -- & -- &  16.1~(4.75) &  14.4~(4.20) \\
\hline
\end{tabular}
\end{center}
\caption{1-$\sigma$ percentage errors on cosmological parameters and isocurvature mode amplitudes for the {\planck}-TE experiment.}
\label{planckte_table}
\end{table*}

\subsection{CMB temperature and E-mode polarization spectrum constraints}
\label{sec4.2}
We now consider the case of {\planck} CMB temperature and E-mode polarization measurements ({\planck}-TE) in constraining the admixture model with the adiabatic and three isocurvature modes. The information from the E-mode polarization spectrum adds significantly to the CMB temperature spectrum constraints, both for cosmological parameters \cite{Galli_etal2014} and for isocurvature mode amplitudes \cite{Bucher_etal2002}. The E-mode polarization measurements also provide a slightly improved reconstruction, in the case of {\planck}, of the lensing potential. \\

The {\planck}-TE forecasts are summarized in Table \ref{planckte_table}. In the case of the pure adiabatic model, adding the E-mode polarization information to the temperature information improves the constraints by up to 35\% for most cosmological parameters. More significantly the large-angle E-mode polarization measurement ($\ell < 30$) breaks the $\tau$-$A_s$ degeneracy, allowing an improvement by a factor of about 7 in constraints on both these parameters. \\

The inclusion of three isocurvature modes in the case of CMB temperature causes the constraints on the cosmological parameters to degrade substantially and the mode amplitudes to be poorly constrained, as noted in Section \ref{sec4.1}. Adding CMB E-mode polarization information significantly improves the cosmological parameter constraints and the isocurvature mode amplitudes are constrained at the 10\% level or better. This improvement results mainly from the large-angle ($\ell < 150$) E-mode polarization measurement, which breaks the degeneracy between the isocurvature mode amplitudes and cosmological parameters \cite{Bucher_etal2002}. The complementary information provided by the large-scale E-mode polarization measurement, as discussed in Section \ref{sec2.2}, can be seen in the bottom left panel of Figure \ref{fig_cancellation_planckte}. There is not much additional information on the most degenerate temperature direction coming from small angular scales in E-mode polarization, as can be seen in the top right panel of Figure \ref{fig_cancellation_planckte}.\\

Adding CMB lensing information to the temperature information, where the lensing field is reconstructed from temperature and E-mode polarization maps, further constrains the admixture model. Constraints on cosmological parameters improve by up to 10\% and constraints on $\tau$ by up to 20\%, similar to the {\planck}-T case. The improvement in constraints on the isocurvature mode amplitudes is again modest, below the 10\% level, and not much improved beyond the {\planck}-T case. This suggests that adding E-mode polarization to the {\planck} lensing reconstruction does not significantly improve the lensing reconstruction error, which is verified by the plot in the lower right panel of Figure \ref{fig_cancellation_planckte}. The improvement arising from CMB lensing is not as drastic as that from E-mode polarization, nevertheless the lensing measurement is useful as it provides a complementary probe of these parameters with different systematics. \\

\begin{table*}[t]
\small
\begin{center}
\begin{tabular}{|l|c|c|c|c|c|c|c|}
\hline
& T & T+E~(+B) & T+L$_{TEB}$~(+B) & T & T+E~(+B) & T+L$_{TEB}$~(+B) & T+E+L$_{TEB}$~(+B) \\
\hline
$\delta \omega_b/\omega_b$&  0.710&  0.272~(0.264) &  0.629~(0.594)&  2.32&  0.324~(0.322) &  1.90~(1.85)&  0.314~(0.313) \\
\hline
$\delta \omega_c/\omega_c$&  1.57&  0.642~(0.572) &  1.26~(0.905)&  5.07&  1.13~(1.05) &  4.39~(2.61)&  0.771~(0.680) \\
\hline
$\delta \Omega_{\Lambda}/\Omega_{\Lambda}$&  1.25&  0.498~(0.443) &  0.993~(0.706)&  3.54&  0.789~(0.726) &  3.07~(1.83)&  0.515~(0.428) \\
\hline
$\delta \tau/\tau$&  40.5&  3.14~(3.13) &  15.5~(11.1)&  47.7&  3.94~(3.87) &  33.0~(29.6)&  3.90~(3.80) \\
\hline
$\delta n_s/n_s$&  0.419&  0.225~(0.215) &  0.391~(0.321)&  1.84&  0.449~(0.438) &  1.70~(1.19)&  0.358~(0.345) \\
\hline
$\delta A_s/A_s$&  6.99&  0.556~(0.553) &  2.33~(1.68)&  30.2&  1.97~(1.16) &  25.5~(5.38)&  1.69~(0.883) \\
\hline
$\langle AD,CI\rangle$& -- & -- & -- &  32.0&  6.13~(6.04) &  29.6~(18.3)&  5.41~(5.39) \\
\hline
$\langle AD,NIV\rangle$& -- & -- & -- &  45.4&  2.51~(1.57) &  43.1~(4.01)&  2.44~(1.44) \\
\hline
$\langle AD,NID\rangle$& -- & -- & -- &  49.1&  4.23~(3.10) &  43.4~(10.4)&  3.59~(2.71) \\
\hline
$\langle CI,CI\rangle$& -- & -- & -- &  32.9&  3.63~(3.63) &  32.2~(17.6)&  3.61~(3.61) \\
\hline
$\langle CI,NIV\rangle$& -- & -- & -- &  20.3&  2.65~(2.56) &  17.8~(11.5)&  2.45~(2.41) \\
\hline
$\langle CI,NID\rangle$& -- & -- & -- &  48.8&  3.92~(3.86) &  47.5~(21.0)&  3.68~(3.66) \\
\hline
$\langle NIV,NIV\rangle$& -- & -- & -- &  17.0&  0.549~(0.474) &  16.7~(4.23)&  0.546~(0.454) \\
\hline
$\langle NIV,NID\rangle$& -- & -- & -- &  36.2&  2.07~(1.46) &  33.9~(7.38)&  1.91~(1.37) \\
\hline
$\langle NID,NID\rangle$& -- & -- & -- &  15.1&  2.09~(1.79) &  12.5~(6.25)&  1.86~(1.63) \\
\hline
\end{tabular}
\end{center}
\caption{1-$\sigma$ percentage errors on cosmological parameters and isocurvature mode amplitudes for the {\prism}-TEB experiment.}
\label{prism_table} 
\end{table*}

\subsection{CMB temperature, E-mode polarization and B-mode polarization spectrum constraints}
\label{sec4.3}

We now study how isocurvature constraints improve when CMB B-mode polarization measurements are added to CMB temperature and E-mode polarization measurements, as in the case of the {\prism}-TEB experiment \cite{PRISM_whitepaper2013}. The improvement in constraining isocurvature perturbations is potentially two-fold: the B-mode polarization measurements provide an enhanced reconstruction of the lensing field, which is sensitive to the primordial perturbations, and the lensing-induced B-mode signal provides an additional probe of the primordial scalar perturbations. \\

The {\prism}-TEB forecasts are summarized in Table \ref{prism_table}. In the case of the pure adiabatic model, adding the lensing-induced B-mode polarization information to the temperature and E-mode polarization information improves the constraints on cosmological parameters by up to 10\%, with marginal improvements on $A_s$ and $\tau$. Adding lensing information improves constraints on all parameters, but most significantly on $\tau$ and $A_s$ by a factor of about 3.\\

\begin{figure}[t!]
\centering
\includegraphics[scale=1.3]{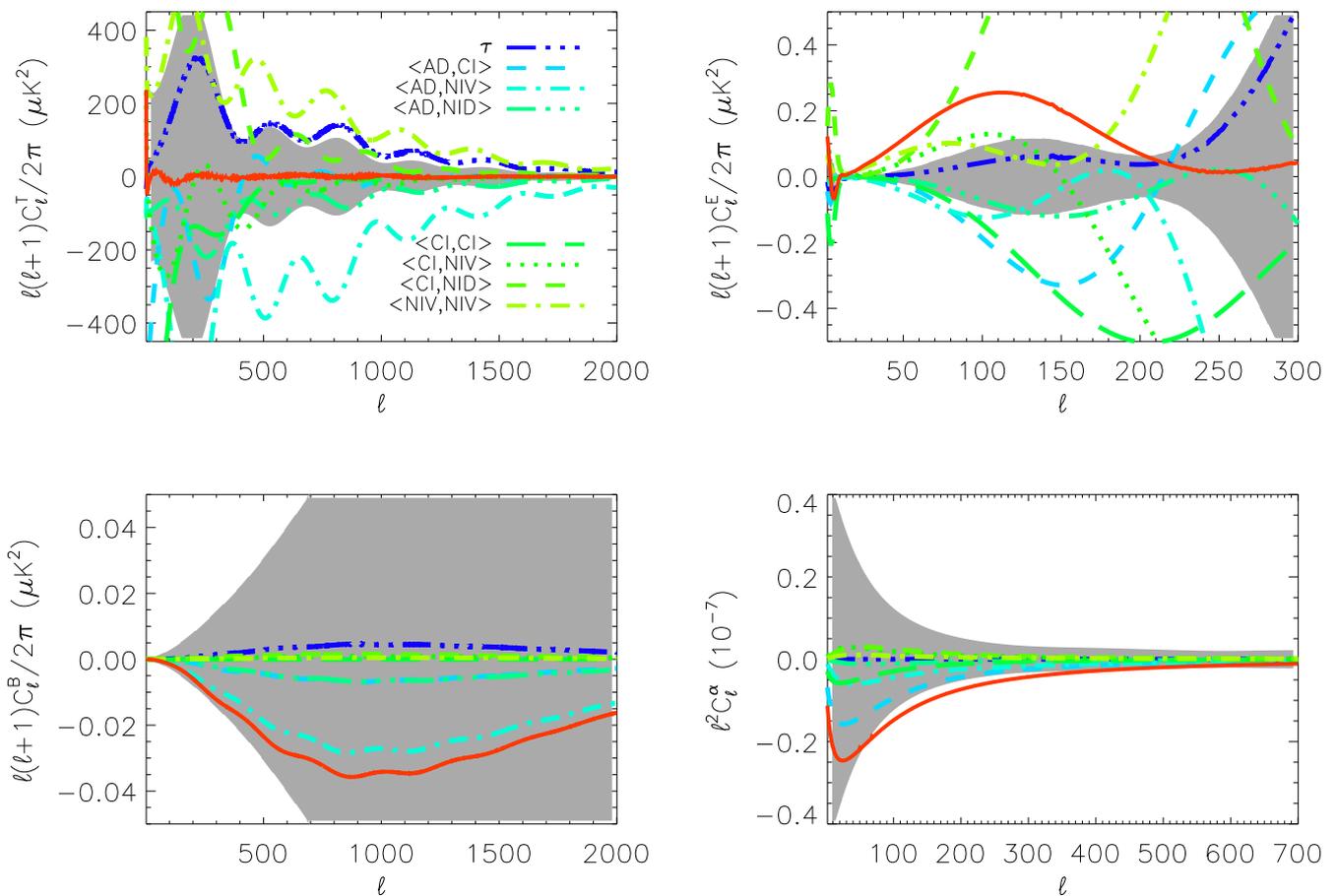}
\vskip -0.5cm 
\caption{Degeneracy breaking.
The top left panel shows the cancellation between cosmological and mode amplitude parameter derivatives using the most degenerate direction of {\prism}-T measurements. The degenerate direction is obtained from a principal component analysis of the Fisher matrix. The red solid line is the sum of all contributions in the degenerate direction and is compared to the power spectrum error represented by the grey shaded region. The top right panel shows how E-mode polarization measurements constrain this degenerate direction. The bottom left and bottom right panels show how lensing-induced B-mode polarization information and lensing information, respectively, add to the CMB temperature degenerate direction.}
\label{fig_cancellation_prismteb}
\end{figure}

For the admixture model, adding lensing-induced B-mode polarization information to the temperature and E-mode polarization information improves the constraints on cosmological parameters by up to 10\%, and on isocurvature mode parameters by up to 40\%. 
It is interesting to consider the flattest direction in the CMB temperature measurement by {\prism} as illustrated in the top left panel of Figure \ref{fig_cancellation_prismteb}. We see from the bottom left panel of Figure \ref{fig_cancellation_prismteb} that there is complementary information in the lensing-induced B-mode polarization spectrum that allows us to constrain isocurvature mode amplitudes. Note that more sensitivity can be gained at a given $\ell$ by binning the spectra. Adding lensing information to the temperature information improves the constraints on isocurvature mode amplitudes by about 10-15\%, and on cosmological parameters by about 20-30\%. The improvement in constraints on isocurvature mode amplitudes from lensing information is illustrated in the bottom right panel of Figure \ref{fig_cancellation_prismteb}. Here it can be seen that the residual lensing spectrum, in the flat direction of the {\prism} temperature measurement, slightly exceeds the lensing reconstruction error over a limited range of angular scales. Adding the combination of lensing and lensing-induced B-mode polarization information to the temperature and E-mode polarization information provides significant improvements on constraints. In particular, isocurvature mode amplitudes are constrained at the few percent level or better.\\

We note that the maximum likelihood lensing estimator \cite{Hirata_Seljak_2003} provides a lower lensing reconstruction noise from the B-mode polarization signal than the quadratic lensing estimator \cite{Hirata_Seljak_2003,PRISM_whitepaper2013,Namikawa_2014}. This could potentially improve the constraints on isocurvature modes. However, we have investigated the effect of including the lower lensing noise for {\prism}-TEB and found that this only improves the constraints very slightly.\\
 
\subsection{Constraints on subsets of isocurvature modes}
\label{sec4.4}

In our main study we considered an admixture model with the adiabatic mode correlated with all three isocurvature modes. In this section, for completeness, we forecast the constraints from the {\planck}-TE and {\prism}-TEB experiments on admixtures of the adiabatic mode with one or two correlated isocurvature modes. \\

For the case of one correlated isocurvature mode, from Table \ref{planck_prism1isomode} we see that the overall constraints on isocurvature mode amplitudes are at the 0.5-1\% level for {\prism}-TEB and at the 1-2\% level for {\planck}-TE. In the case of {\prism} there is a 20-40\% improvement in some isocurvature mode amplitudes when adding the lensing and lensing-induced B-mode polarization information to the temperature and E-mode polarization information. While our results cannot be compared directly to the analysis in \cite{Santos_etal2012}, due to their different parametrizations and choice of parameters, we note that their constraint on $\langle CI,CI\rangle$ of around 1.5\% is similar to ours.\\

For the case of two correlated isocurvature modes, from Table \ref{planck_prism2isomodes} we see that the overall constraints on isocurvature mode amplitudes are at the 1-3\% level for {\prism}-TEB and at the 7\% level or better for {\planck}-TE. For {\prism} the improvement from adding the lensing and lensing-induced B-mode polarization information to the temperature and E-mode polarization information in the case of two correlated isocurvature modes is more significant than in the single isocurvature mode case, with up to 2-3 times better constraints on isocurvature mode amplitudes.

\begin{table*}[t]
\begin{center}
\begin{tabular}{l|cc|ccc}
\hline\hline
&\multicolumn{2}{c|}{{\planck}-TE}& \multicolumn{3}{c}{{\prism}-TEB}\\
\hline
& T~(+L$_{TE}$) & T+E~(+L$_{TE}$) & T~(+L$_{TEB}$) & T+E~(+L$_{TEB}$) & T+E+B~(+L$_{TEB}$)\\
\hline
$\delta A_s/A_s$&  9.03~(4.01) &  1.11~(1.02) &  8.00~(2.44) &  0.600~(0.478) &  0.593~(0.472) \\
$\langle AD,CI\rangle$&  4.56~(4.53) &  2.10~(2.07) &  4.30~(4.27) &  1.09~(0.988) &  1.08~(0.987) \\
$\langle CI,CI\rangle$&  2.19~(1.94) &  1.18~(1.17) &  2.07~(1.84) &  0.536~(0.533) &  0.536~(0.533) \\
\hline
$\delta A_s/A_s$&  8.25~(4.18) &  1.32~(1.29) &  7.47~(2.66) &  0.708~(0.654) &  0.700~(0.611) \\
$\langle AD,NIV\rangle$&  2.82~(2.81) &  1.52~(1.51) &  2.56~(2.55) &  0.692~(0.688) &  0.622~(0.619) \\
$\langle NIV,NIV\rangle$&  1.35~(1.32) &  0.584~(0.583) &  1.25~(1.22) &  0.209~(0.208) &  0.206~(0.205) \\
\hline
$\delta A_s/A_s$&  8.43~(4.10) &  1.23~(1.12) &  7.49~(2.45) &  0.681~(0.515) &  0.680~(0.502) \\
$\langle AD,NID\rangle$&  3.47~(3.46) &  1.53~(1.50) &  3.14~(2.96) &  0.757~(0.689) &  0.755~(0.669) \\
$\langle NID,NID\rangle$&  1.61~(1.53) &  0.677~(0.675) &  1.39~(1.34) &  0.243~(0.243) &  0.241~(0.240) \\
\hline\hline
\end{tabular}
\end{center}
\caption{1-$\sigma$ percentage errors on isocurvature mode amplitudes for the {\planck}-TE and {\prism}-TEB experiments
in the case of an admixture model with the adiabatic mode correlated with one isocurvature mode.}
\label{planck_prism1isomode}
\end{table*} 

\begin{table}[t]
\begin{center}
\begin{tabular}{l|cc|ccc}
\hline\hline
&\multicolumn{2}{c|}{{\planck}-TE}& \multicolumn{3}{c}{{\prism}-TEB}\\
\hline
& T~(+L$_{TE}$) & T+E~(+L$_{TE}$) & T~(+L$_{TEB}$) & T+E~(+L$_{TEB}$) & T+E+B~(+L$_{TEB}$)\\
\hline
$\delta A_s/A_s$&  11.5~(9.83) &  2.66~(2.60) &  10.2~(8.07) &  1.35~(1.16) &  1.08~(0.754) \\
$\langle AD,CI\rangle$&  23.7~(22.4) &  6.60~(6.59) &  20.4~(19.4) &  3.50~(3.32) &  2.53~(2.15) \\
$\langle AD,NIV\rangle$&  13.3~(12.7) &  4.34~(4.34)&  10.9~(10.5) &  2.00~(1.94) &  1.27~(1.10) \\
$\langle CI,CI\rangle$&  9.27~(8.37) &  2.67~(2.66)&  7.94~(7.22) &  1.58~(1.57) &  1.20~(1.17) \\
$\langle CI,NIV\rangle$&  11.6~(10.9) &  3.40~(3.40)&  10.1~(9.33) &  1.57~(1.56) &  1.10~(1.07) \\
$\langle NIV,NIV\rangle$&  4.03~(3.94) &  1.31~(1.31)&  3.44~(3.34) &  0.461~(0.459) &  0.345~(0.337) \\
\hline
$\delta A_s/A_s$&  9.76~(4.77) &  1.74~(1.60)&  8.37~(3.38) &  0.970~(0.862) &  0.920~(0.770) \\
$\langle AD,CI\rangle$&  10.8~(10.5) &  6.13~(6.00)&  9.30~(7.80) &  3.43~(3.06) &  2.88~(2.68) \\
$\langle AD,NID\rangle$&  7.18~(7.01) &  4.05~(3.90)&  5.92~(4.99) &  2.15~(1.94) &  1.82~(1.64) \\
$\langle CI,CI\rangle$&  12.2~(12.0) &  5.32~(5.32)&  10.8~(10.4) &  3.31~(3.28) &  3.30~(3.27) \\
$\langle CI,NID\rangle$&  17.3~(17.2) &  7.19~(7.14)&  15.2~(14.4) &  3.55~(3.47) &  3.55~(3.47) \\
$\langle NID,NID\rangle$&  4.89~(4.85) &  2.44~(2.42)&  3.90~(3.76) &  0.965~(0.927) &  0.965~(0.924) \\
\hline
$\delta A_s/A_s$&  19.3~(11.7) &  4.12~(3.72)&  18.5~(6.73) &  1.90~(1.63) &  1.14~(0.786) \\
$\langle AD,NIV\rangle$&  25.3~(21.9) &  5.41~(5.13)&  24.3~(15.3) &  2.23~(2.12) &  1.01~(0.926) \\
$\langle AD,NID\rangle$&  29.5~(26.0) &  5.43~(5.11)&  28.0~(16.8) &  2.58~(2.40) &  1.31~(1.13) \\
$\langle NIV,NIV\rangle$&  7.80~(6.98) &  1.53~(1.49)&  7.53~(5.52) &  0.466~(0.460) &  0.315~(0.313) \\
$\langle NIV,NID\rangle$&  18.7~(16.8) &  3.12~(3.01)&  18.0~(12.1) &  1.22~(1.18) &  0.620~(0.615) \\
$\langle NID,NID\rangle$&  10.3~(9.48) &  1.72~(1.69)&  9.72~(6.37) &  0.849~(0.828) &  0.528~(0.524) \\
\hline\hline
\end{tabular}
\end{center}
\caption{1-$\sigma$ percentage errors on isocurvature mode amplitudes for the {\planck}-TE and {\prism}-TEB experiments
in the case of an admixture model with the adiabatic mode correlated with two isocurvature modes.}
\label{planck_prism2isomodes}
\end{table} 

\subsection{Sensitivity to model assumptions}
\label{sec4.5}

In this study, we have made a few assumptions that could impact our results. We investigate here the dependence of our results to these assumptions for the case of an admixture model with the adiabatic mode and three isocurvature modes.

\subsubsection{Simultaneous dark energy constraints}

In considering a $\Lambda$CDM cosmology, we have assumed an equation of state, $w=-1,$ that is constant in time. While the $\Lambda$CDM model adequately describes CMB temperature and polarization measurements \cite{Planck_22_2013,Komatsu_etal2011}, other late-time cosmological experiments are able to probe models with evolution in the dark energy component \cite{Amanullah_etal2010,Astier_etal2006}. 
The dark energy parameters affect the growth of matter perturbations and in turn the CMB lensing spectrum, with these variations potentially being degenerate with isocurvature mode amplitudes.\\

We consider a dark energy model with an equation of state modelled by the CPL parameterization, $w(a)=w_0 + (1-a)w_a$, where $a(t)=1/(1 + z)$ is the cosmic scale factor \cite{Chevallier_Polarski_2001,Linder_2003}. We marginalise over cosmological parameters, including the dark energy parameters, to obtain the constraints on isocurvature mode amplitudes. We find that the isocurvature amplitude constraints degrade by at most 10\% for both the {\planck}-TE and {\prism}-TEB experiments when the dark energy parameters are included. This indicates that there is sufficient information in the CMB temperature, polarization and lensing data to distinguish isocurvature perturbations from dark energy.

\subsubsection{CMB temperature-lensing correlation}

In our study we have treated the CMB lensing spectrum as independent of the CMB temperature and polarization spectrum, even though the lensing field is reconstructed from the temperature and polarization maps. Schmittfull {\it et al.} \cite{Schmittfull_etal2013} have shown that this is a reasonable assumption in the case of the CMB temperature-lensing correlation for the {\planck} experiment. For more sensitive experiments, they found that the relevant correlations are most important at low $\ell$s ($\ell^L<150$). \\

To study the effect of overlapping information coming from the CMB temperature and CMB lensing spectra, we implement a cut-off at $\ell_{min}^L=200$ to discard any potential CMB temperature-lensing correlation in the Fisher matrix analysis and study the resulting constraints for the {\prism}-TEB experiment. The effect is small with a degradation in constraints on cosmological parameters and isocurvature mode amplitudes of at most a few percent. This indicates that our main results are insensitive to the information contained in the CMB temperature-lensing correlation. However, further work is required to model the CMB polarization-lensing correlation and study its effect on the constraints presented here.

\subsubsection{Primordial tensor perturbations}

In considering scalar adiabatic and isocurvature perturbations, we have neglected the possibility of a tensor mode contribution, for example, from primordial gravitational waves, for which there has been a recent claimed detection \cite{Ade_etal2014}. If a tensor mode contribution to the B-mode polarization spectrum was present, its effect would be to dilute the information available on large angular scales ($\ell <100$) to constrain the lensing-induced B-mode polarization signal. We implement a cut-off at $\ell_{min}^B=100$ in the Fisher matrix analysis to ignore information from the large-scale CMB B-mode polarization signal in constraining the isocurvature mode amplitudes. We find that there is a negligible increase, of less than 1\%, in the isocurvature mode amplitudes for the {\prism}-TEB experiment when the large-scale B-mode polarization information is not included. This indicates that scalar adiabatic and isocurvature perturbations may be simultaneously constrained with primordial tensor perturbations.

\section{Discussion}
\label{sec5}

In this paper, we have considered the question of how much information is added to CMB temperature and E-mode polarization 
constraints on primordial isocurvature perturbations from CMB lensing information. The CMB lensing information comes from both the CMB lensing spectrum and the lensing-induced B-mode polarization spectrum. In the case of the {\planck} experiment we have found that the CMB lensing spectrum improves constraints by at most 10\%. The limited improvement is a result of the low amplitude of isocurvature lensing spectra relative to the adiabatic lensing spectrum and cancellations between isocurvature modes rendering them only slightly detectable with CMB lensing. The same is true for the lensing spectrum measured by {\prism} even though the lensing reconstruction error is smaller. In the case of the {\prism} experiment there is a much larger improvement in isocurvature constraints from the lensing-induced B-mode polarization spectrum, with mode amplitude constraints improved by up to 40\%. The addition of lensing and lensing-induced B-mode polarization information provides significant improvements, with isocurvature mode amplitudes overall constrained at the few percent level or better.\\

We investigated the sensitivity of our results to various assumptions in our analysis, such as the inclusion of dark energy parameters, the CMB temperature-lensing correlation, and the presence of primordial tensor modes. We found that these assumptions did not significantly change
our main results. Nevertheless, further study of the CMB polarization-lensing correlation is required to understand its impact on the results presented here. Other ways in which our study could be extended include generalising our admixture model to include variations in 
isocurvature spectral indices or investigating how other cosmological probes of the primordial perturbations, for example, galaxy surveys \cite{SDSS-III_overview2011,DES_whitepaper2005,EUCLID_whitepaper2012,LSST_whitepaper2012}, fare in constraining the admixture model. More immediately, though, we aim to use CMB temperature, polarization and lensing datasets to set updated constraints on admixture models with correlated adiabatic and isocurvature modes.

\appendix
\label{sec6}

\acknowledgments
The authors acknowledge support from the National Research Foundation, South Africa. KM acknowledges hospitality of the KITP where part of this work was completed. This research was supported in part by the National Science Foundation (USA) under Grant No. PHY11-25915.\\

\bibliographystyle{simon_bib_default}
\bibliography{my_references}

\end{document}

%% file: newcomm.tex
\newcommand{\gtsim}{\mbox{{\raisebox{-0.4ex}{$\stackrel{>}{{\scriptstyle\sim}}$}}}}
\newcommand{\ltsim}{\mbox{{\raisebox{-0.4ex}{$\stackrel{<}{{\scriptstyle\sim}}$}}}}

\def\gtorder{\mathrel{\raise.3ex\hbox{$>$}\mkern-14mu
             \lower0.6ex\hbox{$\sim$}}}

\newcommand{\ber}{\begin{eqnarray}}
\newcommand{\eer}{\end{eqnarray}}

\newcommand{\ba}{\begin{eqnarray}}
\newcommand{\ea}{\end{eqnarray}}

\newcommand{\planck}{{\sc Planck}} 
 
\newcommand{\prism}{{\sc Prism}}
\newcommand{\core}{{\sc COrE}}
\newcommand{\pixie}{{\sc Pixie}}

%% file: CMB_lensing_forecasts_v5.bbl
\begin{thebibliography}{100}
\providecommand{\url}[1]{\texttt{#1}}
\providecommand{\urlprefix}{URL }
\providecommand{\eprint}[2][]{\url{#2}}

\bibitem{ACBAR_MISSION_2003}
M.~C. {Runyan} \textit{et~al.}, \apjs \textbf{149}, 265 (2003),
  \eprint{astro-ph/0303515}.

\bibitem{ACT_MISSION_2003}
A.~{Kosowsky}, New Astronomy Reviews \textbf{47}, 939 (2003),
  \eprint{astro-ph/0402234}.

\bibitem{ARCHEOPS_MISSION_2004}
M.~{Tristram} \& {Archeops Collaboration}, M.~{Plionis}, ed.,
  \textit{Astrophysics and Space Science Library}, vol. 301 of
  \textit{Astrophysics and Space Science Library}, 97 (2004),
  \eprint{astro-ph/0309349}.

\bibitem{BICEP2_MISSION_2014}
{BICEP2 Collaboration} \textit{et~al.}, ArXiv e-prints  (2014),
  \eprint{1403.4302}.

\bibitem{BOOMERANG_MISSION_2006}
C.~J. {MacTavish} \textit{et~al.}, \apj \textbf{647}, 799 (2006),
  \eprint{astro-ph/0507503}.

\bibitem{CBI_MISSION_2001}
S.~{Padin} \textit{et~al.}, \apjl \textbf{549}, L1 (2001),
  \eprint{astro-ph/0012211}.

\bibitem{COBE_MISSION_1990}
J.~C. {Mather} \textit{et~al.}, \apjl \textbf{354}, L37 (1990).

\bibitem{MAXIMA_MISSION_2006}
B.~{Rabii} \textit{et~al.}, Review of Scientific Instruments \textbf{77},
  071101 (2006), \eprint{astro-ph/0309414}.

\bibitem{PLANCK_MISSION_2014}
F.~R. {Bouchet} \& {on behalf of the Planck collaboration for the results},
  ArXiv e-prints  (2014), \eprint{1405.0439}.

\bibitem{QMAP_MAT_TOCO_MISSIONS_2002}
A.~{Miller} \textit{et~al.}, \apjs \textbf{140}, 115 (2002),
  \eprint{astro-ph/0108030}.

\bibitem{SPT_MISSION_2004}
J.~{Ruhl} \textit{et~al.}, C.~M. {Bradford} \textit{et~al.}, eds.,
  \textit{Z-Spec: a broadband millimeter-wave grating spectrometer: design,
  construction, and first cryogenic measurements}, vol. 5498 of \textit{Society
  of Photo-Optical Instrumentation Engineers (SPIE) Conference Series}, 11--29
  (2004), \eprint{astro-ph/0411122}.

\bibitem{WMAP_MISSION_2003}
C.~L. {Bennett} \textit{et~al.}, \apj \textbf{583}, 1 (2003),
  \eprint{astro-ph/0301158}.

\bibitem{CAPMAP_first_results_2003}
D.~{Barkats}, New Astronomy Reviews \textbf{47}, 1077 (2003),
  \eprint{astro-ph/0306002}.

\bibitem{DASI_MISSION_2002}
E.~M. {Leitch} \textit{et~al.}, \apj \textbf{568}, 28 (2002),
  \eprint{astro-ph/0104488}.

\bibitem{MAXIPOL_MISSION_2003}
B.~R. {Johnson} \textit{et~al.}, New Astronomy Reviews \textbf{47}, 1067
  (2003), \eprint{astro-ph/0308259}.

\bibitem{PIQUE_MISSION_1997}
E.~J. {Wollack} \textit{et~al.}, \apj \textbf{476}, 440 (1997),
  \eprint{astro-ph/9601196}.

\bibitem{QUAD_first_results_2008}
P.~{Ade} \textit{et~al.}, \apj \textbf{674}, 22 (2008), \eprint{0705.2359}.

\bibitem{QUIET_MISSION_2012}
{QUIET Collaboration} \textit{et~al.}, ArXiv e-prints  (2012),
  \eprint{1207.5562}.

\bibitem{VSA_first_results_2003}
R.~A. {Watson} \textit{et~al.}, \mnras \textbf{341}, 1057 (2003),
  \eprint{astro-ph/0205378}.

\bibitem{Story_etal2013}
K.~T. {Story} \textit{et~al.}, \apj \textbf{779}, 86 (2013),
  \eprint{1210.7231}.

\bibitem{Bennett_etal2013}
C.~L. {Bennett} \textit{et~al.}, \apjs \textbf{208}, 20 (2013),
  \eprint{1212.5225}.

\bibitem{Hinshaw_etal2013}
G.~{Hinshaw} \textit{et~al.}, \apjs \textbf{208}, 19 (2013),
  \eprint{1212.5226}.

\bibitem{Planck_22_2013}
{Planck Collaboration} \textit{et~al.}, ArXiv e-prints  (2013),
  \eprint{1303.5082}.

\bibitem{Hou_etal2014}
Z.~{Hou} \textit{et~al.}, \apj \textbf{782}, 74 (2014), \eprint{1212.6267}.

\bibitem{Das_etal2014}
S.~{Das} \textit{et~al.}, \jcap \textbf{4}, 014 (2014), \eprint{1301.1037}.

\bibitem{Ade_etal2014}
A.~R. Ade, P.\ \textit{et~al.}, \prl \textbf{112}, 241101 (2014).

\bibitem{Essinger-Hileman_etal2010}
T.~{Essinger-Hileman} \textit{et~al.}, ArXiv e-prints  (2010),
  \eprint{1008.3915}.

\bibitem{Filippini_etal2010}
J.~P. {Filippini} \textit{et~al.}, \textit{Society of Photo-Optical
  Instrumentation Engineers (SPIE) Conference Series}, vol. 7741 of
  \textit{Society of Photo-Optical Instrumentation Engineers (SPIE) Conference
  Series} (2010).

\bibitem{Austermann_etal2012}
J.~E. {Austermann} \textit{et~al.}, \textit{Society of Photo-Optical
  Instrumentation Engineers (SPIE) Conference Series}, vol. 8452 of
  \textit{Society of Photo-Optical Instrumentation Engineers (SPIE) Conference
  Series} (2012), \eprint{1210.4970}.

\bibitem{Niemack_etal2010}
M.~D. {Niemack} \textit{et~al.}, \textit{Society of Photo-Optical
  Instrumentation Engineers (SPIE) Conference Series}, vol. 7741 of
  \textit{Society of Photo-Optical Instrumentation Engineers (SPIE) Conference
  Series} (2010).

\bibitem{POLARBEAR_etal2014}
{The POLARBEAR Collaboration} \textit{et~al.}, ArXiv e-prints  (2014),
  \eprint{1403.2369}.

\bibitem{Naess_etal2014}
S.~{Naess} \textit{et~al.}, ArXiv e-prints  (2014), \eprint{1405.5524}.

\bibitem{Sievers_etal2013}
J.~L. {Sievers} \textit{et~al.}, \jcap \textbf{10}, 060 (2013),
  \eprint{1301.0824}.

\bibitem{Bucher_etal2000}
M.~{Bucher}, K.~{Moodley} \& N.~{Turok}, \prd \textbf{62}, 083508 (2000),
  \eprint{astro-ph/9904231}.

\bibitem{Mollerach_1990}
S.~{Mollerach}, Phys.Lett. \textbf{B242}, 158 (1990).

\bibitem{Polarski_Starobinsky_1994}
D.~{Polarski} \& A.~A. {Starobinsky}, \prd \textbf{50}, 6123 (1994),
  \eprint{astro-ph/9404061}.

\bibitem{GarciaBellido_Wands_1996}
J.~{Garc{\'{\i}}a-Bellido} \& D.~{Wands}, \prd \textbf{53}, 5437 (1996),
  \eprint{astro-ph/9511029}.

\bibitem{Linde_Mukhanov_1997}
A.~{Linde} \& V.~{Mukhanov}, \prd \textbf{56}, 535 (1997),
  \eprint{astro-ph/9610219}.

\bibitem{Langlois_1999}
D.~{Langlois}, \prd \textbf{59}, 123512 (1999), \eprint{astro-ph/9906080}.

\bibitem{Gordon_etal2001}
C.~{Gordon} \textit{et~al.}, \prd \textbf{63}, 023506 (2001),
  \eprint{astro-ph/0009131}.

\bibitem{Langlois_2008}
D.~{Langlois}, Journal of Physics Conference Series \textbf{140}, 012004
  (2008), \eprint{0809.2540}.

\bibitem{Efstathiou_Bond_1986}
G.~{Efstathiou} \& J.~R. {Bond}, \mnras \textbf{218}, 103 (1986).

\bibitem{Bozza_etal2002}
V.~{Bozza} \textit{et~al.}, Physics Letters B \textbf{543}, 14 (2002),
  \eprint{hep-ph/0206131}.

\bibitem{Lyth_Wands_2002}
D.~H. {Lyth} \& D.~{Wands}, Physics Letters B \textbf{524}, 5 (2002),
  \eprint{hep-ph/0110002}.

\bibitem{Moroi_Takahashi_2001}
T.~{Moroi} \& T.~{Takahashi}, Physics Letters B \textbf{522}, 215 (2001),
  \eprint{hep-ph/0110096}.

\bibitem{Bartolo_Liddle_2002}
N.~{Bartolo} \& A.~R. {Liddle}, \prd \textbf{65}, 121301 (2002),
  \eprint{astro-ph/0203076}.

\bibitem{Moroi_Takahashi_2002}
T.~{Moroi} \& T.~{Takahashi}, \prd \textbf{66}, 063501 (2002),
  \eprint{hep-ph/0206026}.

\bibitem{Lyth_etal2003}
D.~H. {Lyth}, C.~{Ungarelli} \& D.~{Wands}, \prd \textbf{67}, 023503 (2003),
  \eprint{astro-ph/0208055}.

\bibitem{Dimopoulos_etal2003a}
K.~{Dimopoulos} \textit{et~al.}, Journal of High Energy Physics \textbf{5}, 057
  (2003), \eprint{hep-ph/0303154}.

\bibitem{Dimopoulos_etal2003b}
K.~{Dimopoulos} \textit{et~al.}, Journal of High Energy Physics \textbf{7}, 053
  (2003), \eprint{hep-ph/0304050}.

\bibitem{Iliesiu_etal2014}
L.~{Iliesiu} \textit{et~al.}, \prd \textbf{89}, 103513 (2014),
  \eprint{1312.3636}.

\bibitem{Brown_etal2009}
M.~L. {Brown} \textit{et~al.}, \apj \textbf{705}, 978 (2009),
  \eprint{0906.1003}.

\bibitem{Reichardt_etal2009}
C.~L. {Reichardt} \textit{et~al.}, \apj \textbf{694}, 1200 (2009),
  \eprint{0801.1491}.

\bibitem{Larson_etal2011}
D.~{Larson} \textit{et~al.}, \apjs \textbf{192}, 16 (2011), \eprint{1001.4635}.

\bibitem{Reid_etal2010}
B.~A. {Reid} \textit{et~al.}, \mnras \textbf{404}, 60 (2010),
  \eprint{0907.1659}.

\bibitem{Percival_etal2010}
W.~J. {Percival} \textit{et~al.}, \mnras \textbf{401}, 2148 (2010),
  \eprint{0907.1660}.

\bibitem{Carbone_etal2011}
C.~{Carbone}, A.~{Mangilli} \& L.~{Verde}, \jcap \textbf{9}, 028 (2011),
  \eprint{1107.1211}.

\bibitem{Amanullah_etal2010}
R.~{Amanullah} \textit{et~al.}, \apj \textbf{716}, 712 (2010),
  \eprint{1004.1711}.

\bibitem{Bucher_etal2001}
M.~{Bucher}, K.~{Moodley} \& N.~{Turok}, \prl \textbf{87}, 191301 (2001),
  \eprint{astro-ph/0012141}.

\bibitem{Peiris_etal2003}
H.~V. {Peiris} \textit{et~al.}, \apjs \textbf{148}, 213 (2003),
  \eprint{astro-ph/0302225}.

\bibitem{Hinshaw_etal2007}
G.~{Hinshaw} \textit{et~al.}, \apjs \textbf{170}, 288 (2007),
  \eprint{astro-ph/0603451}.

\bibitem{Komatsu_etal2011}
E.~{Komatsu} \textit{et~al.}, \apjs \textbf{192}, 18 (2011),
  \eprint{1001.4538}.

\bibitem{Enqvist_etal2000}
K.~{Enqvist}, H.~{Kurki-Suonio} \& J.~{V{\"a}liviita}, \prd \textbf{62}, 103003
  (2000), \eprint{astro-ph/0006429}.

\bibitem{Andrade_etal2005}
A.~P. {Andrade}, C.~A. {Wuensche} \& A.~L. {Ribeiro}, \prd \textbf{71}, 043501
  (2005), \eprint{astro-ph/0501399}.

\bibitem{Sievers_etal2007}
J.~L. {Sievers} \textit{et~al.}, \apj \textbf{660}, 976 (2007),
  \eprint{astro-ph/0509203}.

\bibitem{Stompor_etal1996}
R.~{Stompor}, A.~J. {Banday} \& K.~M. {Gorski}, \apj \textbf{463}, 8 (1996),
  \eprint{astro-ph/9511087}.

\bibitem{Langlois_Riazuelo_2000}
D.~{Langlois} \& A.~{Riazuelo}, \prd \textbf{62}, 043504 (2000),
  \eprint{astro-ph/9912497}.

\bibitem{Amendola_etal2002}
L.~{Amendola} \textit{et~al.}, \prl \textbf{88}, 211302 (2002),
  \eprint{astro-ph/0107089}.

\bibitem{Dawson_etal2013}
K.~S. {Dawson} \textit{et~al.}, \aj \textbf{145}, 10 (2013),
  \eprint{1208.0022}.

\bibitem{Tegmark_etal2004}
M.~{Tegmark} \textit{et~al.}, \apj \textbf{606}, 702 (2004),
  \eprint{astro-ph/0310725}.

\bibitem{Samushia_etal2013}
L.~{Samushia} \textit{et~al.}, \mnras \textbf{429}, 1514 (2013),
  \eprint{1206.5309}.

\bibitem{Astier_etal2006}
P.~{Astier} \textit{et~al.}, \aap \textbf{447}, 31 (2006),
  \eprint{astro-ph/0510447}.

\bibitem{Valiviita_Muhonen_2003}
J.~{V{\"a}liviita} \& V.~{Muhonen}, \prl \textbf{91}, 131302 (2003),
  \eprint{astro-ph/0304175}.

\bibitem{Gordon_Malik_2004}
C.~{Gordon} \& K.~A. {Malik}, \prd \textbf{69}, 063508 (2004),
  \eprint{astro-ph/0311102}.

\bibitem{Moodley_etal2004}
K.~{Moodley} \textit{et~al.}, \prd \textbf{70}, 103520 (2004),
  \eprint{astro-ph/0407304}.

\bibitem{Bean_etal2006}
R.~{Bean}, J.~{Dunkley} \& E.~{Pierpaoli}, \prd \textbf{74}, 063503 (2006),
  \eprint{astro-ph/0606685}.

\bibitem{Savelainen_etal2013}
M.~Savelainen \textit{et~al.}, \prd \textbf{88}, 063010 (2013).

\bibitem{Crotty_etal2003}
P.~Crotty \textit{et~al.}, \prl \textbf{91}, 171301 (2003).

\bibitem{Beltran_etal2004}
M.~Beltr\'an \textit{et~al.}, \prd \textbf{70}, 103530 (2004).

\bibitem{Kurki-Suonio_etal2005}
H.~{Kurki-Suonio}, V.~{Muhonen} \& J.~{V{\"a}liviita}, \prd \textbf{71}, 063005
  (2005), \eprint{astro-ph/0412439}.

\bibitem{Beltran_etal2005}
M.~{Beltr{\'a}n} \textit{et~al.}, \prd \textbf{72}, 103515 (2005),
  \eprint{astro-ph/0509209}.

\bibitem{Bucher_etal2004}
M.~{Bucher} \textit{et~al.}, \prl \textbf{93}, 081301 (2004),
  \eprint{astro-ph/0401417}.

\bibitem{Dunkley_etal2005}
J.~Dunkley \textit{et~al.}, \prl \textbf{95}, 261303 (2005).

\bibitem{Planck_17_2013}
{Planck Collaboration} \textit{et~al.}, ArXiv e-prints  (2013),
  \eprint{1303.5077}.

\bibitem{Santos_etal2012}
L.~{Santos} \textit{et~al.}, \prd \textbf{86}, 023002 (2012),
  \eprint{1206.2832}.

\bibitem{CAMB_1999}
A.~{Lewis}, A.~{Challinor} \& A.~{Lasenby}, \apj \textbf{538}, 473 (2000),
  \eprint{astro-ph/9911177}.

\bibitem{Bucher_etal2002}
M.~{Bucher}, K.~{Moodley} \& N.~{Turok}, \prd \textbf{66}, 023528 (2002),
  \eprint{astro-ph/0007360}.

\bibitem{Hu_Sugiyama_1994}
W.~{Hu} \& N.~{Sugiyama}, \apj \textbf{436}, 456 (1994),
  \eprint{astro-ph/9403031}.

\bibitem{Lewis_etal2000}
A.~{Lewis}, A.~{Challinor} \& A.~{Lasenby}, \apj \textbf{538}, 473 (2000),
  \eprint{astro-ph/9911177}.

\bibitem{Trotta_PhD2004}
R.~{Trotta}, \textit{{Cosmic Microwave Background Anisotropies: Beyond Standard
  Parameters}}, Ph.D. thesis, PhD Thesis, 2004 (2004).

\bibitem{Giovannini_2005}
M.~{Giovannini}, International Journal of Modern Physics D \textbf{14}, 363
  (2005), \eprint{astro-ph/0412601}.

\bibitem{MuyaKasanda_etal2012}
S.~{Muya Kasanda} \textit{et~al.}, \jcap \textbf{7}, 021 (2012),
  \eprint{1111.2572}.

\bibitem{Hu_Sugiyama_1995a}
W.~{Hu} \& N.~{Sugiyama}, \prd \textbf{51}, 2599 (1995),
  \eprint{astro-ph/9411008}.

\bibitem{Zaldarriaga_2004}
M.~{Zaldarriaga}, Measuring and Modeling the Universe , 309 (2004),
  \eprint{astro-ph/0305272}.

\bibitem{Galli_etal2014}
S.~{Galli} \textit{et~al.}, ArXiv e-prints  (2014), \eprint{1403.5271}.

\bibitem{Moodley_MuyaKasanda_2013}
S.~{Muya Kasanda} \& K.~{Moodley}, {\it in prep.} (2014).

\bibitem{Kodama_Sasaki_1986}
H.~Kodama \& M.~Sasaki, Int.J.Mod.Phys. \textbf{A1}, 265 (1986).

\bibitem{Cabella_Kamionkowski_2004}
P.~{Cabella} \& M.~{Kamionkowski}, ArXiv e-prints  (2004),
  \eprint{astro-ph/0403392}.

\bibitem{Baumann_etal2009}
D.~{Baumann} \textit{et~al.}, S.~{Dodelson} \textit{et~al.}, eds.,
  \textit{American Institute of Physics Conference Series}, vol. 1141 of
  \textit{American Institute of Physics Conference Series}, 10--120 (2009),
  \eprint{0811.3919}.

\bibitem{Lewis_Challinor_2006}
A.~{Lewis} \& A.~{Challinor}, \physrep \textbf{429}, 1 (2006),
  \eprint{astro-ph/0601594}.

\bibitem{Tegmark_etal1997}
M.~{Tegmark}, A.~N. {Taylor} \& A.~F. {Heavens}, \apj \textbf{480}, 22 (1997),
  \eprint{astro-ph/9603021}.

\bibitem{Eisenstein_etal1999}
D.~J. {Eisenstein}, W.~{Hu} \& M.~{Tegmark}, \apj \textbf{518}, 2 (1999),
  \eprint{astro-ph/9807130}.

\bibitem{Schmittfull_etal2013}
M.~M. {Schmittfull} \textit{et~al.}, \prd \textbf{88}, 063012 (2013),
  \eprint{1308.0286}.

\bibitem{Perotto_etal_2006}
L.~{Perotto} \textit{et~al.}, \jcap \textbf{10}, 013 (2006),
  \eprint{astro-ph/0606227}.

\bibitem{PLANCK_MISSION_2006}
{The Planck Collaboration}, ArXiv Astrophysics e-prints  (2006),
  \eprint{astro-ph/0604069}.

\bibitem{PRISM_whitepaper2013}
{PRISM Collaboration} \textit{et~al.}, ArXiv e-prints  (2013),
  \eprint{1306.2259}.

\bibitem{PIXIE_whitepaper2011}
A.~{Kogut} \textit{et~al.}, \jcap \textbf{7}, 025 (2011), \eprint{1105.2044}.

\bibitem{CORE_whitepaper2011}
{The COrE Collaboration} \textit{et~al.}, ArXiv e-prints  (2011),
  \eprint{1102.2181}.

\bibitem{Albrecht_etal2006}
A.~{Albrecht} \textit{et~al.}, ArXiv e-prints  (2006),
  \eprint{astro-ph/0609591}.

\bibitem{Hu_Okamoto_2002}
W.~{Hu} \& T.~{Okamoto}, \apj \textbf{574}, 566 (2002),
  \eprint{astro-ph/0111606}.

\bibitem{Planck_1_2013}
{Planck Collaboration} \textit{et~al.}, ArXiv e-prints  (2013),
  \eprint{1303.5062}.

\bibitem{Hirata_Seljak_2003}
C.~M. {Hirata} \& U.~{Seljak}, \prd \textbf{68}, 083002 (2003),
  \eprint{astro-ph/0306354}.

\bibitem{Namikawa_2014}
T.~{Namikawa}, Progress of Theoretical and Experimental Physics \textbf{2014},
  06B108 (2014), \eprint{1403.3569}.

\bibitem{Chevallier_Polarski_2001}
M.~{Chevallier} \& D.~{Polarski}, \ijmpd \textbf{10}, 213 (2001),
  \eprint{gr-qc/0009008}.

\bibitem{Linder_2003}
E.~V. Linder, \prl \textbf{90}, 091301 (2003).

\bibitem{SDSS-III_overview2011}
D.~J. {Eisenstein} \textit{et~al.}, \aj \textbf{142}, 72 (2011),
  \eprint{1101.1529}.

\bibitem{DES_whitepaper2005}
{The Dark Energy Survey Collaboration}, ArXiv Astrophysics e-prints  (2005),
  \eprint{astro-ph/0510346}.

\bibitem{EUCLID_whitepaper2012}
J.~{Amiaux} \textit{et~al.}, \textit{Society of Photo-Optical Instrumentation
  Engineers (SPIE) Conference Series}, vol. 8442 of \textit{Society of
  Photo-Optical Instrumentation Engineers (SPIE) Conference Series} (2012),
  \eprint{1209.2228}.

\bibitem{LSST_whitepaper2012}
{LSST Dark Energy Science Collaboration}, ArXiv e-prints  (2012),
  \eprint{1211.0310}.

\end{thebibliography}
